\documentclass[a4paper,11pt]{article}

\usepackage{jheppub}

\usepackage{graphicx}
\usepackage{mathrsfs}
\usepackage{amsmath}
\usepackage{amssymb}
\usepackage{bm}
\usepackage{bbm}
\usepackage{color}
\usepackage{slashed}
\usepackage{soul}

\allowdisplaybreaks[4]

\title{Scale-invariant total decay width $\Gamma(H\to b\bar{b})$ using the novel method of characteristic operator}

\author[a]{Jiang Yan,}
\emailAdd{yjiang@cqu.edu.cn}

\author[a]{Xing-Gang Wu,}
\emailAdd{wuxg@cqu.edu.cn}

\author[b]{Jian-Ming Shen,}
\emailAdd{shenjm@hnu.edu.cn}

\author[c]{Xu-Dong Huang,}
\emailAdd{huangxd@cqnu.edu.cn}

\author[a]{Zhi-Fei Wu}
\emailAdd{wuzf@cqu.edu.cn}

\affiliation[a]{Department of Physics, Chongqing Key Laboratory for Strongly Coupled Physics, Chongqing University, Chongqing 401331, P.R. China}
\affiliation[b]{School of Physics and Electronics, Hunan Provincial Key Laboratory of High-Energy Scale Physics and Applications, Hunan University, Changsha 410082, P.R. China}
\affiliation[c]{College of Physics and Electronic Engineering, Chongqing Normal University, Chongqing 401331, P.R. China}

\abstract{
According to the renormalization group invariance, any physical observable should be invariant under different renormalization schemes and scales. Nevertheless, a fixed-order pQCD approximant cannot inherently satisfy this requirement, giving rise to the conventional scheme-and-scale ambiguities. The principle of maximum conformality (PMC) offers a process-independent approach to address these two types of ambiguities. In practice, it is necessary to handle the $n_{f}$-terms in the pQCD series with caution to obtain accurate PMC predictions. In this paper, a novel method via using the characteristic operator~(CO) ${\cal \hat{D}}_{n_{\gamma}, n_{\beta}}$ is proposed to extend the applicability of PMC, which is a theoretical generalization of previous PMC single-scale setting approach. The CO framework not only streamlines derivations for complex scenarios, yielding simplified procedures and more compact expressions, but also achieves a scheme-and-scale invariant pQCD series by fixing the correct effective magnitude of $\alpha_s$ and the running mass simultaneously. Both are well matched with the expansion coefficients of the series, leading to the wanted scheme-and-scale invariant conformal series. As an example, we show the achievement of scale-invariant N$^{4}$LO total decay width $\Gamma(H\to b\bar{b})$ under the $\overline{\rm MS}$-scheme. Using the CO framework, its effective coupling $\alpha_{s}(Q_{*})$ and effective $b$-quark $\overline{\rm MS}$-mass $\overline{m}_{b}(Q_{*})$ are determined by absorbing all non-conformal $\{\beta_{i}\}$-terms from the renormalization group equations for either $\alpha_s$ or $\overline{m}_{b}$ simultaneously. The PMC scale is fixed up to N$^3$LL-accuracy, $Q_{*} = 55.2916$~GeV and a scale-invariant total decay width is obtained, $\Gamma(H \to b\bar{b}) = 2.3819 _{-0.0231}^{+0.0230}$~MeV, whose errors are squared averages of the ones associated with $\Delta \alpha_{s}(M_{Z}) = \pm 0.0009$, $\Delta M_{H} = 0.11$~GeV, $\Delta \overline{m}_{b}(\overline{m}_{b}) = \pm 0.007$~GeV, and the uncalculated N$^{5}$LO contributions $\Delta\Gamma= \pm0.0001$~MeV predicted via Bayesian analysis with the degree-of-belief ${\rm DoB}=95.5\%$.
}

\begin{document}

\maketitle

\flushbottom

\section{Introduction} \label{sec1}

Due to asymptotic freedom property of Quantum Chromodynamics (QCD)~\cite{Gross:1973id, Politzer:1973fx}, the strong running coupling ($\alpha_s$) becomes numerically small at short distances, allowing perturbative calculations of cross sections or decay widths for the high momentum transfer processes. To regularize the ultra-violet divergences encountered in perturbative higher-order calculations, a specific renormalization scheme and scale must be introduced. According to the requirement of renormalization group invariance~(RGI)~\cite{StueckelbergdeBreidenbach:1952pwl, Peterman:1978tb, Callan:1970yg, Symanzik:1970rt}, a valid prediction for a physical observable must be independent of the choices of renormalization scheme and scale. However a truncated perturbation series does not automatically satisfy these requirements, since the renormalization scheme/scale dependence from both the strong running coupling and the perturbative QCD (pQCD) calculable coefficients cannot be exactly cancelled for an arbitrary choice of scale. In practice, the renormalization scale $\mu_{r}$ of a pQCD series is usually chosen as $Q$, which is the scale to remove the large logs of the series or the one to match the measured value of the observable, and then vary it within a range, such as $\mu_{r}\in [Q/\xi, \xi Q]$, with $\xi$ being selected as $2$, $3$, $4$, or etc., to account for a rough order estimation of its uncertainties. This method is referred to as conventional scale-setting approach. This approach is evidently arbitrary, and the convergent behavior of the pQCD series depends heavily on the choice of $\mu_r$. Moreover, due to the divergent renormalon terms like $n! \alpha_{s}^{n} \beta_{0}^{n}$ emerged in the initial pQCD series are generally proportional to some powers of the log terms $\ln\mu_r^2/Q^2$~\footnote{This property explains why conventional series under a specific choice of $\mu_r=\eta Q$ has similar perturbative behavior of the PMC series, where the specific value of $\eta$ can be determined by equaling the conventional result to be the same as the PMC one.}, the net scale error of the series will become larger for the choice of a broader scale range with larger $\xi$. The conventional scale error of the initial series is usually treated as a systematic error of the pQCD prediction. Thereby the conventional scale-setting approach significantly diminishes the predictive capability of pQCD theory. It is thus important to have a proper scale-setting approach, so as to achieve a more precise fixed-order pQCD series that satisfies the standard RGI~\cite{Wu:2013ei, Wu:2014iba}.

Many scale-setting approaches have been suggested in the literature. Among them, the principle of maximum conformality~(PMC)~\cite{Brodsky:2011ta, Brodsky:2011ig, Mojaza:2012mf, Brodsky:2012rj, Brodsky:2013vpa}, being  a successful extension of the well-known Brodsky-Lepage-Mackenzie~(BLM) method~\cite{Brodsky:1982gc}, provides a process independent and systematic way to eliminate the conventional scheme-and-scale errors. The BLM method suggests to deal with the $n_f$-terms of the series. Since the running behavior of $\alpha_{s}$ is governed by the renormalization group equation~(RGE) or the $\beta$ function~\cite{Gross:1973ju, Politzer:1974fr}, the PMC method suggests that to handle the RGE-involved $\{\beta_{i}\}$-terms of the series is more fundamental than to deal with the $n_{f}$-terms. Practically, the PMC method unambiguously translates the $n_{f}$-terms associated with the RGE into the $\{\beta_{i}\}$-terms by employing the general QCD degeneracy relations among different orders~\cite{Bi:2015wea}, and subsequently determines the PMC scales at each order by absorbing the relevant $\{\beta_{i}\}$-terms. It is noted that all the features previously observed in the BLM literature are also adaptable to PMC with or without proper transformations~\footnote{Most importantly, one needs to confirm that whether the $n_f$-terms have been correctly treated in previous BLM predictions.}. Notably, the BLM and the PMC naturally converge to the well-known Gell-Mann-Low method~\cite{Gell-Mann:1954yli} for QED in the Abelian limit~\cite{Brodsky:1997jk}.

Lately, the PMC single-scale-setting method~(PMCs), being an effective alternative to the original PMC multi-scale-setting method (PMCm), was proposed in Refs.\cite{Shen:2017pdu, Yan:2022foz} from two distinct but equivalent perspectives. Using the PMCs, it has been demonstrated that the PMC prediction is independent of the choices of renormalization scheme and scale~\cite{Wu:2018cmb}, agreeing well with the self-consistency requirements of the renormalization group~\cite{Brodsky:2012ms, Wu:2019mky}. Calculations are often performed most advantageously in the modified minimal-subtraction scheme ($\overline{\rm MS}$-scheme), but all references to such theoretically constructed scheme may be eliminated when comparisons are made between observables. This also avoids the problem that one need not expand observables in terms of couplings which have singular or ill-defined functional dependence~\cite{Grunberg:1980ja, Grunberg:1982fw, Grunberg:1989xf}. Additionally, the convergence of PMC series is often significantly enhanced due to the absorption of divergent renormalon terms~\cite{Beneke:1994qe, Neubert:1994vb, Beneke:1998ui} into $\alpha_{s}(Q_{*})$. Moreover, the PMCs approach can effectively reduce the residual scale dependence~\cite{Zheng:2013uja} of the PMC predictions, which are due to uncalculated terms or unknown higher-order~(UHO) terms in the pQCD series. To date, PMC has been successfully applied in numerous high-energy physics studies under the scale-invariant mass scheme, examples can be found in the reviews~\cite{Wu:2013ei, Wu:2019mky, DiGiustino:2023jiq}, in which one needs to correctly deal with $\{\beta_i\}$-terms that govern $\alpha_s$ scale-running behavior via RGE~\footnote{Following the QCD factorization theory, the QCD prediction of a physical observable can be factorized into perturbatively calculable hard part and the non-perturbative part such as the parton distribution function or the fragmentation function or the matrix element or the wavefunction at the zero, and etc. So the PMC should be applied with great care so as to achieve a precise prediction free of renormalization scheme-and-scale ambiguities, especially it is important to derive the full renormalization and factorization scale dependent terms before applying the PMC.}. Practically, to avoid the misuse of $n_f$-terms to fix the magnitude of $\alpha_s$, one usually transforms the $\overline{\rm MS}$-mass series to the pole-mass one. Recently, the PMC has been utilized to determine the top-quark and bottom-quark on-shell masses via their perturbative relations to the $\overline{\rm MS}$-scheme quark masses up to four-loop QCD corrections, marking the first successful application of PMC to deal with the pQCD series with the $\overline{\rm MS}$-scheme quark mass terms. For the purpose, new degeneracy relations have been elucidated with the aid of RGEs incorporating both the $\beta$-function and the quark mass anomalous dimension $\gamma_{m}$~\cite{Huang:2022rij, Ma:2024xeq}.

As a fundamental particle predicted by the Standard Model~(SM) of particle physics, the Higgs boson~\cite{Higgs:1964ia, Englert:1964et, Higgs:1964pj, Guralnik:1964eu} was discovered in 2012 by the ATLAS and CMS Collaborations at the Large Hadron Collider~(LHC)~\cite{ATLAS:2012yve, CMS:2012qbp}. The substantial Yukawa coupling of the Higgs boson to the top and bottom quarks significantly impacts the SM predictions of the physical observables. Thus, the Higgs boson provides an outstanding platform for investigating fundamental interactions at and beyond the electroweak symmetry breaking scale. Around the electroweak symmetry breaking scale, the decay channel $H\to b\bar{b}$ is the predominant decay mode of the SM Higgs boson, which constitutes about $(53\pm 8)\%$ of all its decay processes~\cite{ATLAS:2022vkf, ParticleDataGroup:2024cfk}. Thus the investigation of $H\to b\bar{b}$ is crucial for precision tests of the SM and for exploring new physics beyond the SM.

Total decay width $\Gamma(H \to b\bar{b})$ is significantly influenced by the QCD corrections. The QCD corrections at ${\cal O}(\alpha_{s})$-level, incorporating the dependence on the bottom quark $\overline{\rm MS}$-scheme running mass $\overline{m}_{b}$, have been examined in Refs.~\cite{Braaten:1980yq, Sakai:1980fa, Inami:1980qp, Drees:1990dq, Kataev:1993be}. Besides the trivial overall factor associated with the Yukawa coupling, QCD corrections up to ${\cal O}(\alpha_{s}^{4})$-level have been computed for the case of massless quarks in Refs.~\cite{Gorishnii:1990zu, Gorishnii:1991zr, Chetyrkin:1996sr, Baikov:2005rw, Chetyrkin:2010dx, Herzog:2017dtz, Chen:2023fba}. Effects of finite top-quark and bottom-quark masses up to ${\cal O}(\alpha_{s}^{2})$-level have been calculated in Ref.~\cite{Larin:1995sq, Harlander:1997xa, Primo:2018zby, Wang:2023xud}. Very recently, the bottom quark mass effect has been calculated at ${\cal O}(\alpha_{s}^{3})$-level~\cite{Wang:2024ilc}. The hadronic decay of Higgs boson, encompassing contributions from both $H\to b\bar{b}$ and $H\to gg$ channels, has been studied up to ${\cal O}(\alpha_{s}^{4})$-level in Refs.~\cite{Chetyrkin:1997vj, Davies:2017xsp, Zeng:2018jzf, Zeng:2020lwi, Abbas:2024mme}, and the analysis based on Pad\'{e} approximants has been done in Ref.~\cite{Boito:2021scm}. Moreover, the next-to-leading order~(NLO) electroweak corrections as well as the mixed QCD and electroweak corrections at the ${\cal O}(\alpha\alpha_{s})$-level have been calculated in Refs.~\cite{Dabelstein:1991ky, Kniehl:1991ze, Kataev:1997cq, Mihaila:2015lwa}. Those works provide us with great chances to achieve precise pQCD prediction on $\Gamma(H \to b\bar{b})$.

At present, the Higgs decay width into $b\bar{b}$ has been analyzed using the PMC multi-scale-setting approach -- PMCm -- in Ref.~\cite{Wang:2013bla} and the PMC single-scale-setting approach -- PMCs -- in Ref.~\cite{Shen:2017pdu}. The scale-invariant predictions have been derived by using either PMCm or PMCs. The resulting PMCm and PMCs predictions are nearly identical for both the total and differential contributions. As mentioned above, the PMCs method suggests to use an overall scale to effectively replace those individual PMC scales derived under the PMCm method in the sense of a mean value theorem~\cite{Shen:2017pdu, Yan:2022foz}. The PMC single scale can be regarded as the overall effective momentum flow of the process; it also shows stability and convergence with increasing order in pQCD via the pQCD approximates. In this paper, we will adopt the PMCs to do our discussions.

To further improve the PMCs scale-setting procedures, we will first introduce a new characteristic operator~(CO) ${\cal \hat{D}}_{n_{\gamma},n_{\beta}}$ to formalize the PMCs procedures and then apply it for $\Gamma(H \to b\bar{b})$. This operator characterizes the scale-running behavior for both $\alpha_s$ and the $\overline{\rm MS}$-scheme quark mass and derive more compact expressions for the new degeneracy relations and the previous PMC formulas. Moreover, given that $\overline{m}_{b} \ll M_{H}$ and the top-quark mass effect starts at ${\cal O}(\alpha_{s}^{2})$-level, the contributions of quark masses are negligible, and to concentrate our attention on the PMC CO framework, the analysis presented in this paper will also be conducted using the massless quarks in the QCD corrections up to N$^{4}$LO level. It is noted that employing the CO method allows these new formulas to be easily extended to the cases with more scale running quantities, if their behavior are governed by the similar RGE-like equations.

For a perturbative series, in addition to the conventional scale uncertainty or the residual scale uncertainty, it is also helpful to have a reliable estimation on the contribution of UHO-terms for the perturbative series. Given the unknown exact pQCD result, it is beneficial to quantify the UHO-terms' contribution using a probability distribution. Following  Bayesian analysis (BA)~\cite{Cacciari:2011ze, Bagnaschi:2014wea, Bonvini:2020xeo, Duhr:2021mfd}, the conditional probability of the unknown coefficient is initially provided by a subjective prior distribution, which is then updated iteratively according to the Bayes' theorem as more information becomes available. It has been found that the generally more convergent and scheme-and-scale invariant PMC series provides a more reliable basis than the initial scale-dependent series for estimating the contributions from the UHO-terms~\cite{Du:2018dma, Shen:2022nyr, Shen:2023qgz, Luo:2023cpa, Yan:2023mjj,Yan:2024hbz}. In this paper, we will adopt the BA approach to estimate the contributions of the not-yet-calculated N$^{5}$LO-terms for the PMC series.

The remaining parts of the paper are organized as follows. In Sec.~\ref{sec2}, we provide a detailed introduction to the PMCs approach based on the characteristic operator. In Sec.~\ref{sec3}, we analyze the total decay width $\Gamma(H\to b\bar{b})$ up to N$^{4}$LO QCD corrections by applying the PMCs. Numerical results and discussions are presented in Sec.~\ref{sec4}. Section~\ref{sec5} is reserved for the summary.

\section{Calculation technology}\label{sec2}

Without loss of generality, we write the pQCD prediction $\rho$ calculated under the $\overline{\rm MS}$-scheme, together with the use of $\overline{\rm MS}$-scheme running quark mass, in the following form
\begin{align}
	\rho\left(\mu_{r},\alpha_{s}(\mu_{r}),\overline{m}_{q}(\mu_{r});Q\right) = \overline{m}_{q}^{n_{\gamma}}(\mu_{r})\alpha_{s}^{n_{\beta}}(\mu_{r})R\left(\mu_{r},\alpha_{s}(\mu_{r}),\overline{m}_{q}(\mu_{r});Q\right),
\end{align}
where $\mu_{r}$ denotes an arbitrary initial choice of renormalization scale, $Q$ signifies the kinetic scale at which the physical observable is measured or the typical momentum flow of the reaction, $\overline{m}_{q}$ denotes the $\overline{\rm MS}$ mass, $n_{\gamma}$ ($n_{\gamma} \ge 0$) is the $\overline{m}_{q}$-order, $n_{\beta}$ ($n_{\beta} \ge 0$, but $n_{\gamma}$ and $n_{\beta}$ can not be $0$ simultaneously\footnote{If $n_{\gamma}=n_{\beta}=0$ at the LO, then we should redefine $\rho$ to be the remainder of the subtracted LO term.}) is the $\alpha_{s}$-order of the leading-order~(LO) contribution. $R\left(\mu_{r},\alpha_{s}(\mu_{r}),\overline{m}_{q}(\mu_{r});Q\right)$ represents the kernel function of the pQCD prediction $\rho$, which can be expanded as a power series over $\alpha_s$, i.e.
\begin{align}
	R\left(\mu_{r},\alpha_{s}(\mu_{r}),\overline{m}_{q}(\mu_{r});Q\right) = \sum_{i=1}^{\infty} r_{i}\left(\mu_{r}/Q,\overline{m}_{q}(\mu_{r})\right)\alpha_{s}^{i-1}(\mu_{r}).
\end{align}
A physical observable must comply with the requirement of RGI, that is,
\begin{align}
	\frac{{\rm d}}{{\rm d} \ln\mu_{r}^{2}}\rho\left(\mu_{r},\alpha_{s}(\mu_{r}),\overline{m}_{q}(\mu_{r});Q\right) = \frac{{\rm d}}{{\rm d} \ln\mu_{r}^{2}}\left[\overline{m}_{q}^{n_{\gamma}}(\mu_{r})\alpha_{s}^{n_{\beta}}(\mu_{r})R\left(\mu_{r},\alpha_{s}(\mu_{r}),\overline{m}_{q}(\mu_{r});Q\right)\right]=0,
\end{align}
which results in
\begin{align}\label{R_eq}
	\left(\frac{\partial}{\partial \ln\mu_{r}^{2}} + \overline{m}_{q}\gamma_{m}\frac{\partial }{\partial \overline{m}_{q}}\right) R = -\left(n_{\gamma}\gamma_{m} + n_{\beta}\frac{\beta}{\alpha_{s}} + \beta\frac{\partial}{\partial\alpha_{s}}\right)R,
\end{align}
where the operator on the left-hand side (l.h.s) of Eq.~\eqref{R_eq} acts only on the coefficients of the kernel function $R$, while the operator on the right-hand side (r.h.s.) acts only on $\alpha_{s}$ within $R$, i.e.,
\begin{align}\label{r_eq}
	\sum_{i=1}^{\infty} \alpha_{s}^{i-1}\left(\frac{\partial}{\partial \ln\mu_{r}^{2}} + \overline{m}_{q}\gamma_{m}\frac{\partial }{\partial \overline{m}_{q}}\right) r_{i} = -\sum_{i=1}^{\infty} r_{i} \left(n_{\gamma}\gamma_{m} + n_{\beta}\frac{\beta}{\alpha_{s}} + \beta\frac{\partial}{\partial\alpha_{s}}\right)\alpha_{s}^{i-1}
\end{align}
Note that Eq.~\eqref{R_eq} or Eq.~\eqref{r_eq} are so written such that the l.h.s. operator acts on the process-dependent coefficients alone, while the r.h.s. operator acts on the $\alpha_{s}$ itself. This separation enables us to investigate the correct scale-running behavior of the kernel function $R$ or the perturbative series by examining the operators' action on $\alpha_{s}$ within the r.h.s. via a process independent way.~\footnote{In fact, observables containing any scale-dependent quantity $\cal O$ satisfying the RGE ${\rm d}{\cal O}(\mu_r)/{\rm d} \ln \mu^{2}_r = {\cal O}(\mu_r) \gamma_{\cal O}(\alpha_{s})$ can be treated by this method and are not limited to the running quark masses.}

\subsection{characteristic operator~(CO)}

Now we can define the characteristic operator~(CO) ${\cal \hat{D}}_{n_{\gamma},n_{\beta}}$ as
\begin{align}\label{COdef}
	{\cal \hat{D}}_{n_{\gamma},n_{\beta}} := n_{\gamma}\gamma_{m} + n_{\beta}\frac{\beta}{\alpha_{s}} + \beta\frac{\partial}{\partial\alpha_{s}}.
\end{align}
Here, the $\beta(\alpha_{s})$-function and the quark mass anomalous dimension $\gamma_{m}(\alpha_{s})$ govern the scale evolution behavior of $\alpha_{s}$ and $\overline{m}_{q}$, respectively, as follows:
\begin{align}
	\frac{{\rm d} \alpha_{s}(\mu)}{{\rm d} \ln \mu^{2}} &= \beta(\alpha_{s}) = -\sum_{i=0}\beta_{i}\alpha_{s}^{i+2}(\mu),\\
	\frac{{\rm d} \overline{m}_{q}(\mu)}{{\rm d} \ln \mu^{2}} &= \overline{m}_{q}(\mu)\gamma_{m}(\alpha_{s}) = -\overline{m}_{q}(\mu) \sum_{i=0}\gamma_{i}\alpha_{s}^{i+1}(\mu).
\end{align}
At present, the $\{\beta_i\}$-functions and $\{\gamma_{i}\}$-functions have been computed up to the five-loop level in the $\overline{\rm MS}$-scheme~\cite{Gross:1973ju, Politzer:1973fx, Gross:1973id, Politzer:1974fr, Chetyrkin:1997dh, Vermaseren:1997fq, Chetyrkin:2004mf, Baikov:2014qja, Baikov:2016tgj, Herzog:2017ohr}, e.g., $\beta_{0} = \left(11 C_{A}/3 - 4 T_{F} n_{f}/3\right)/(4\pi)$ and $\gamma_{0} = 3 C_{F}/(4\pi)$ for $n_{f}$ active flavors, where $C_{A} = 3$, $C_{F} = 4/3$ and $T_{F} = 1/2$ are SU(3)$_{\rm C}$ color factors. In particular, we can restate the more general case of the above formula in terms of the CO as
\begin{align}
	\frac{{\rm d}^{k} \alpha_{s}^{n_{\beta}}(\mu)}{\left({\rm d} \ln \mu^{2}\right)^{k}} &= \alpha_{s}^{n_{\beta}}(\mu) \left(n_{\beta}\frac{\beta}{\alpha_{s}} + \beta\frac{\partial}{\partial\alpha_{s}}\right)^{k}[1] = \alpha_{s}^{n_{\beta}}(\mu) {\cal \hat{D}}_{0,n_{\beta}}^{k}[1],\label{da}\\
	\frac{{\rm d}^{k} \overline{m}_{q}^{n_{\gamma}}(\mu)}{\left({\rm d} \ln \mu^{2}\right)^{k}} &= \overline{m}_{q}^{n_{\gamma}}(\mu) \left(n_{\gamma}\gamma_{m} + \beta\frac{\partial}{\partial\alpha_{s}}\right)^{k}[1]= \overline{m}_{q}^{n_{\gamma}}(\mu) {\cal \hat{D}}_{n_{\gamma},0}^{k}[1],\label{dm}
\end{align}
and
\begin{align}
	\frac{{\rm d}^{k} \left[\overline{m}_{q}^{n_{\gamma}}(\mu)\alpha_{s}^{n_{\beta}}(\mu)\right]}{\left({\rm d} \ln \mu^{2}\right)^{k}} = \overline{m}_{q}^{n_{\gamma}}(\mu) \alpha_{s}^{n_{\beta}}(\mu) {\cal \hat{D}}_{n_{\gamma},n_{\beta}}^{k}[1]\label{dma},
\end{align}
where the symbol ``$[1]$'' represents that the preceding operator acts on $\alpha_s^{0}=1$. From Eqs.~\eqref{da}-\eqref{dma}, we have the following simple relation
\begin{align}
	{\cal \hat{D}}_{n_{\gamma},n_{\beta}}^{k}[1] = \sum_{i=0}^{k}C_{k}^{i}{\cal \hat{D}}_{n_{\gamma},0}^{i}[1]{\cal \hat{D}}_{0,n_{\beta}}^{k-i}[1],
\end{align}
where $C_{j}^{k}=j!/\left(k!(j-k)!\right)$ are binomial coefficients. This simple relation allows us to differentiate the contributions of various running quantities within the pQCD series.

From the definition~\eqref{COdef}, the CO only includes the differentiation of the coupling $\alpha_{s}$, thus we only need to consider the properties of the CO operates on $\alpha_{s}^{\ell}$, that is, ${\cal \hat{D}}_{n_{\gamma},n_{\beta}}^{k}[\alpha_{s}^{\ell}]$. According to Eq.~\eqref{COdef}, it is straightforward to obtain the following expression:
\begin{align}\label{series_Das}
	{\cal \hat{D}}_{n_{\gamma},n_{\beta}}^{k}[\alpha_{s}^{\ell}] = (-1)^{k}\sum_{i=0}^{\infty} d_{i}^{[n_{\gamma},n_{\beta};k,\ell]}\alpha_{s}^{\ell+k+i},
\end{align}
where $d_{i}^{[n_{\gamma},n_{\beta};k,\ell]}$ is a series of coefficients containing only $\{\beta_{i}\}$-functions or $\{\gamma_{i}\}$-functions. The detailed expressions of $d_{i}^{[n_{\gamma},n_{\beta};k,\ell]}$ are shown in Appendix~\ref{A1}. We can also find that
\begin{align}
	d_{i}^{[n_{\gamma},n_{\beta}+\lambda;k,\ell]}=d_{i}^{[n_{\gamma},n_{\beta};k,\ell+\lambda]},
\end{align}
which means
\begin{align}\label{asD}
	\alpha_{s}^{\lambda}{\cal \hat{D}}_{n_{\gamma},n_{\beta}+\lambda}^{k}[\alpha_{s}^{\ell}]={\cal \hat{D}}_{n_{\gamma},n_{\beta}}^{k}[\alpha_{s}^{\ell+\lambda}].
\end{align}
Here, we can re-derive the scale-displacement relation of the couplings, running mass, or the combination of these at two different scales $\mu_{1}$ and $\mu_{2}$,
\begin{align}\label{SDR}
	\overline{m}_{q}^{n_{\gamma}}(\mu_{1}) \alpha_{s}^{n_{\beta}}(\mu_{1}) & = \overline{m}_{q}^{n_{\gamma}}(\mu_{2}) \alpha_{s}^{n_{\beta}}(\mu_{2}) + \sum_{k=1}^{\infty} \frac{(-1)^{k}}{k!}\frac{{\rm d}^{k} \left[\overline{m}_{q}^{n_{\gamma}}(\mu)\alpha_{s}^{n_{\beta}}(\mu)\right]}{\left({\rm d} \ln \mu^{2}\right)^{k}} \bigg|_{\mu=\mu_{2}} \ln^{k}\frac{\mu_{2}^{2}}{\mu_{1}^{2}}\notag\\
	& = \overline{m}_{q}^{n_{\gamma}}(\mu_{2}) \alpha_{s}^{n_{\beta}}(\mu_{2})\left[ 1 + \sum_{k=1}^{\infty} \frac{(-1)^{k}}{k!} \ln^{k}\frac{\mu_{2}^{2}}{\mu_{1}^{2}} {\cal \hat{D}}_{n_{\gamma},n_{\beta}}^{k}[1] \right]\notag\\
	& = \overline{m}_{q}^{n_{\gamma}}(\mu_{2}) \alpha_{s}^{n_{\beta}}(\mu_{2})\left[ 1 + \sum_{\ell=1}^{\infty} \sum_{k=1}^{\ell} \frac{1}{k!} \ln^{k}\frac{\mu_{2}^{2}}{\mu_{1}^{2}} d_{\ell-k}^{[n_{\gamma},n_{\beta};k,0]} \alpha_{s}^{\ell} \right].
\end{align}
In particular, the general scale-displacement relation for $\alpha_{s}$ or $\overline{m}_{q}$ can be obtained by setting $n_{\gamma}=0$ or $n_{\beta}=0$ in the above formulas, respectively.

\subsection{The principle of maximum conformality~(PMC)}

As mentioned in the Introduction, to concentrate our attention on the PMC CO framework, the analysis presented in this paper will be conducted using the massless quarks in the QCD corrections up to N$^{4}$LO level. Then its kernel function $R$ will not explicitly contain the $\overline{m}_{q}$-terms. The initial series of $R$ is usually given in $n_f$-series with $n_f$ being the active number of quark flavors. To applying the PMC correctly, one need to apply the $\beta(\alpha_{s})$-function and the quark mass anomalous dimension $\gamma_{m}(\alpha_{s})$ to reexpress $R$ as a $\{\beta_i\}$ and $\gamma_{j}$ series so as to achieve the correct magnitudes of $\alpha_s$ and $\bar{m}_q$ simultaneously. For the purpose, we need to derive new degeneracy relations among different orders. The degeneracy relations introduced by the PMC are required by the conformality of the final series~\cite{Bi:2015wea}; they show that the $\beta$-pattern for the pQCD series at each order is a superposition of the $\{\beta_i\}$-terms which govern the evolution of all of the lower-order $\alpha_s$ contributions. Conversely, they determine the correct running behavior of $\alpha_s$ of the considered process. The new degeneracy relations can be obtained by employing Eqs.~(\ref{asD}, \ref{SDR}), e.g. the required new general QCD degeneracy relations are
\begin{align}
	r_{1}(\mu_{r}/Q) =&\, r_{1,0},\label{DR1}\\
	r_{2}(\mu_{r}/Q) =&\, r_{2,0} + d_{0}^{[n_{\gamma},n_{\beta};1,0]} r_{2,1},\label{DR2}\\
	r_{3}(\mu_{r}/Q) =&\, r_{3,0} + d_{1}^{[n_{\gamma},n_{\beta};1,0]} r_{2,1} + d_{0}^{[n_{\gamma},n_{\beta};1,1]} r_{3,1} + \frac{1}{2!} d_{0}^{[n_{\gamma},n_{\beta};2,0]} r_{3,2},\label{DR3}\\
	r_{4}(\mu_{r}/Q) =&\, r_{4,0} + d_{2}^{[n_{\gamma},n_{\beta};1,0]} r_{2,1} + d_{1}^{[n_{\gamma},n_{\beta};1,1]} r_{3,1} + d_{0}^{[n_{\gamma},n_{\beta};1,2]} r_{4,1}\notag\\
	&\, + \frac{1}{2!} \left( d_{1}^{[n_{\gamma},n_{\beta};2,0]} r_{3,2} + d_{0}^{[n_{\gamma},n_{\beta};2,1]} r_{4,2} \right) + \frac{1}{3!} d_{0}^{[n_{\gamma},n_{\beta};3,0]} r_{4,3},\label{DR4}\\
	r_{5}(\mu_{r}/Q) =&\, r_{5,0} + d_{3}^{[n_{\gamma},n_{\beta};1,0]} r_{2,1} +d_{2}^{[n_{\gamma},n_{\beta};1,1]} r_{3,1} + d_{1}^{[n_{\gamma},n_{\beta};1,2]} r_{4,1} + d_{0}^{[n_{\gamma},n_{\beta};1,3]} r_{5,1}\notag\\
	&\, + \frac{1}{2!} \left( d_{2}^{[n_{\gamma},n_{\beta};2,0]} r_{3,2} + d_{1}^{[n_{\gamma},n_{\beta};2,1]} r_{4,2} + d_{0}^{[n_{\gamma},n_{\beta};2,2]} r_{5,2} \right)\notag\\
	&\, + \frac{1}{3!} \left( d_{1}^{[n_{\gamma},n_{\beta};3,0]} r_{4,3} + d_{0}^{[n_{\gamma},n_{\beta};3,1]} r_{5,3} \right) + \frac{1}{4!} d_{0}^{[n_{\gamma},n_{\beta};4,0]} r_{5,4},\label{DR5}
\end{align}
where the scale-dependent perturbative coefficients $r_{i,j}$ can be redefined as
\begin{align}
	r_{i,j} = \sum_{k=0}^{j} C_{j}^{k} \hat{r}_{i-k,j-k}\ln^{k}\frac{\mu_{r}^{2}}{Q^{2}},
\end{align}
where $\hat{r}_{i,j} = r_{i,j}|_{\mu_{r}=Q}$. Specifically, $\hat{r}_{i,0} = r_{i,0}$ are scale-independent conformal coefficients. Finally, the kernel function $R$ can be rewritten in a more compact form:
\begin{align}\label{R_DR}
	R(\mu_{r},\alpha_{s}(\mu_{r});Q) =&\, \sum_{i=1}^{\infty} r_{i,0}\alpha_{s}^{i-1}(\mu_{r})\notag\\
	&\, + \sum_{i=2}^{\infty} \sum_{j=1}^{i-1} \frac{(-1)^{j}}{j!} \sum_{k=0}^{j} C_{j}^{k} \hat{r}_{i-k,j-k}\ln^{k}\frac{\mu_{r}^{2}}{Q^{2}} {\cal \hat{D}}_{n_{\gamma},n_{\beta}}^{j} \left[\alpha_{s}^{i-j-1}(\mu_{r})\right].
\end{align}

Following the standard PMCs procedure, all scale-dependent non-conformal terms should be absorbed into $\overline{m}_{q}(Q_{*})$ and $\alpha_{s}(Q_{*})$, where $Q_{*}$ is a global effective scale and its value can be determined by requiring all non-conformal terms to vanish in Eq.~\eqref{R_DR}. For a fixed-order pQCD approximant of the kernel function $R$, such as up to $\alpha_{s}^{n-1}$-order, one can thus determine the PMC scale $Q_{*}$ by solving the following equation,
\begin{align}\label{PMC_scale_Eq}
	\sum_{i=2}^{n} \sum_{j=1}^{i-1} \frac{(-1)^{j}}{j!} \sum_{k=0}^{j} C_{j}^{k} \hat{r}_{i-k,j-k}\ln^{k}\frac{Q_{*}^{2}}{Q^{2}} {\cal \hat{D}}_{n_{\gamma},n_{\beta}}^{j} \left[\alpha_{s}^{i-j-1}(Q_{*})\right]=0,
\end{align}
where $n \ge 2$ indicates that the PMC scale $Q_{*}$ can be fixed up to N$^{n-2}$LL accuracy. Equation \eqref{PMC_scale_Eq} can be solved numerically or asymptotically. The asymptotic solution $Q_{*,(\rm asy)}$ is expressed in the form of $\ln \left(Q_{*}^{2}/Q^{2}\right)$ by expanding \eqref{PMC_scale_Eq} in terms of $\alpha_{s}(Q_{*})$ or $\alpha_{s}(Q_{0})$ (where $Q_{0}$ is typically selected as $Q$) up to the required N$^{n-2}$LL accuracy~\cite{Shen:2017pdu, Yan:2022foz}. The relationship between those two expansions has been elaborated in Ref.~\cite{Yan:2022foz}. Here we adopt the following asymptotic solution to do our discussion,
\begin{align}\label{Qs}
	\ln\frac{Q_{*}^{2}}{Q^{2}} = \sum_{i=0}^{n-2} S_{i} \alpha_{s}^{i}(Q_{*}),
\end{align}
where the expressions of coefficients $S_{i}$ are given in Appendix~\ref{A2} for convenience.

\section{The Higgs boson inclusive decay channel $H \to b\bar{b}$}
\label{sec3}

The total decay width of the Higgs boson decays into a $b\bar{b}$ pair is given by
\begin{align}
	\Gamma(H \to b\bar{b}) = \frac{G_{F}M_{H}}{4\sqrt{2}\pi} \overline{m}_{b}^{2}(M_{H}) \tilde{R}(s = M_{H}^{2}),
\end{align}
where $\tilde{R}(\mu;s) = {\rm Im}\,\tilde{\Pi}(-s-i\epsilon)/(2\pi s)$ represents the absorptive part of the scalar two-point correlator $\tilde{\Pi}(\mu;Q)$~\cite{Baikov:2005rw, Chetyrkin:2010dx} which satisfies the following RGE~\cite{Sakai:1980fa, Chetyrkin:1996sr} when $\tilde{\Pi}(\mu;Q)$ is massless:
\begin{align}\label{RGE_Pi}
	\left(\frac{\partial}{\partial\ln\mu^{2}}+\beta(\alpha_{s})\frac{\partial}{\partial\alpha_{s}}+2\gamma^{\rm S}(\alpha_{s})\right) \tilde{\Pi}(\mu;Q) = Q^{2} \gamma^{\rm SS}(\alpha_{s}),
\end{align}
where
\begin{equation}
	\frac{\tilde{\Pi}(\mu;Q)}{Q^{2}} = N_{C}\sum_{k=0}^{\infty} \Pi_{k}(\mu/Q) \alpha_{s}^{k}(\mu)
\end{equation}
and
\begin{equation}
	\gamma^{\rm SS} = N_{C}\sum_{k=0}^{\infty} \gamma^{\rm SS}_{k} \alpha_{s}^{k}(\mu),
\end{equation}
which are expansions of the Higgs polarization operator and the anomalous dimension, respectively, which have been calculated up to ${\cal O}(\alpha_{s}^{4})$-level accuracy~\cite{Baikov:2005rw, Chetyrkin:2010dx}. Additionally, $\gamma^{\rm S}(\alpha_{s}) = \gamma_{m}(\alpha_{s})$ when considering only the lowest order in Yukawa coupling. For uniformity, the analytic expressions for $\gamma^{\rm SS}_{k}$ and $\Pi_{k}$ are listed in Appendix~\ref{A3}.

The Adler function $\tilde{D}(\mu;Q)$, defined in the Euclidean region, has been introduced to associate the observable defined in the Minkowskian space as
\begin{align}
	\tilde{D}(\mu;Q) &= \frac{1}{6}\frac{{\rm d}}{{\rm d}\ln Q^{2}}\frac{\tilde{\Pi}(\mu;Q)}{Q^{2}} = \int_{0}^{\infty} {\rm d}s \frac{Q^{2}\tilde{R}(\mu;s)}{\left(s+Q^{2}\right)^{2}}, \label{Adler1} \\
	\tilde{R}(\mu;s) &= N_{C}\sum_{i=1}^{\infty} r_{i}(\mu^{2}/s) \alpha_{s}^{i-1}(\mu),\label{Conv_Hbb}
\end{align}
where $r_{1}=1$ and $r_{i}$ at arbitrary scale $\mu$ are only functions of $\ln \left(\mu^{2}/s\right)$. Since the $\ln Q^{2}$ terms in $\tilde{\Pi}(\mu;Q)/Q^{2}$ always appear as $\ln \left(\mu^{2}/Q^{2}\right)$, $\tilde{D}$ can also be written as follows according to Eq.\eqref{RGE_Pi},
\begin{align} \label{Adler2}
	\tilde{D}(\mu;Q) &= -\frac{1}{6}\frac{\partial}{\partial\ln \mu^{2}}\frac{\tilde{\Pi}(\mu;Q)}{Q^{2}} = \frac{1}{6}\left( 2\gamma_{m}(\alpha_{s}) + \beta(\alpha_{s})\frac{\partial}{\partial\alpha_{s}} \right) \frac{\tilde{\Pi}(\mu;Q)}{Q^{2}} - \frac{1}{6}\gamma^{\rm SS}(\alpha_{s})\notag\\
	&= \sum_{i=1}^{\infty} \tilde{d}_{i}(\mu/Q) \alpha_{s}^{i-1}(\mu),
\end{align}
then the relations between $r_{i}$ and $\tilde{d}_{i}$ can be obtained by using the elementary integrals
\begin{align}
	&\,\int_{0}^{\infty} {\rm d}s\,\frac{Q^{2}}{\left(s+Q^{2}\right)^{2}}\left\{1; \ln\frac{\mu^{2}}{s}; \ln^{2}\frac{\mu^{2}}{s}; \ln^{3}\frac{\mu^{2}}{s}; \ln^{4}\frac{\mu^{2}}{s}\right\}\notag\\
	=&\, \left\{1; \ln\frac{\mu^{2}}{Q^{2}}; \ln^{2}\frac{\mu^{2}}{Q^{2}} + \frac{\pi^{2}}{3}; \ln^{3}\frac{\mu^{2}}{Q^{2}} + \pi^{2} \ln\frac{\mu^{2}}{Q^{2}}; \ln^{4}\frac{\mu^{2}}{Q^{2}} + 2\pi^{2}\ln^{2}\frac{\mu^{2}}{Q^{2}} + \frac{7\pi^{4}}{15} \right\}.
\end{align}
As expected, the resulting expressions for the coefficients of $\tilde{R}$ have exactly the form of Eqs.~(\ref{DR1})-(\ref{DR5}), with the following identification of the coefficients $r_{i,j}$:
\begin{align}
	r_{i,0}   &= -\frac{\gamma^{\rm SS}_{i-1}}{2},&
	r_{i+1,1} &= -\frac{\Pi_{i-1}}{2},&
	r_{i+2,2} &= \frac{\pi^{2}}{6} \gamma^{\rm SS}_{i-1},&
	r_{i+3,3} &= \frac{\pi^{2}}{2} \Pi_{i-1},&
	r_{i+4,4} &= - \frac{\pi^{4}}{10} \gamma^{\rm SS}_{i-1},
\end{align}
where $i \ge 1$. In Ref.~\cite{AlamKhan:2023dms}, the kinematic terms that are proportional to $\pi^{2}$ have been further studied. It is noted that the $\gamma^{\rm SS}_{k}$ contain $n_{f}$-terms, but since they are not RGE-involved to any order, they must be kept fixed in the PMC scale-setting procedure. It has been repeatedly stressed in previous PMC works that only the $n_{f}$-terms associated with the running of $\alpha_{s}$ should be transformed into $\{\beta_{i}\}$-terms to accurately determine the value of $\alpha_{s}$, and by doing so, correct PMC predictions can be achieved. For example, by using the RGI as a guiding principle, Refs.~\cite{Wu:2019mky, Shen:2016dnq, Ma:2015dxa} explain why the expansion coefficients of the QED anomalous dimension, which appears in the definition of the Adler function, must be treated as conformal coefficients. We put more explanation in the Appendix~\ref{A4} for self-consistency. The condition is similar for the present case.

Following the standard PMC procedure, we are able to achieve a scheme-and-scale independent conformal series for $\Gamma(H \to b\bar{b})$ up to N$^{4}$LO QCD corrections, e.g.
\begin{align}\label{PMC_Hbb}
	\Gamma(H \to b\bar{b})|_{\rm PMC} = \frac{G_{F}M_{H}}{4\sqrt{2}\pi} \overline{m}_{b}^{2}(Q_{*}) N_{C} \sum_{i=1}^{5} r_{i,0} \alpha_{s}^{i-1}(Q_{*}).
\end{align}

\section{Numerical results and discussions}
\label{sec4}

For numerical calculations, the following values from the Particle Data Group~\cite{ParticleDataGroup:2024cfk} are taken as the input parameters,
\begin{align}
	\overline{m}_{b}(\overline{m}_{b})&=4.183\pm 0.007\ {\rm GeV},&M_{H}&=125.20 \pm 0.11\ {\rm GeV},\notag\\
	M_{Z}&=91.1876\ {\rm GeV},&G_{F}&=1.1663788\times10^{-5}\ {\rm GeV}^{2}, \notag
\end{align}
and $\alpha_{s}(M_{Z})=0.1180\pm 0.0009$.

\subsection{Basic properties}

Using Eq.~\eqref{PMC_scale_Eq}, we obtain the desired LL-, NLL-, N$^{2}$LL and N$^{3}$LL-accuracy PMC scales that correspond to NLO, N$^{2}$LO, N$^{3}$LO and N$^{4}$LO pQCD series, respectively, e.g.,
\begin{align}\label{Qs_ext}
	Q_{*}^{(\rm LL, NLL, N^{2}LL, N^{3}LL)}=\{46.0585, 52.9381, 55.1600, 55.2916\}\ {\rm GeV},
\end{align}
or equivalently, using Eqs.~(\ref{S0})-(\ref{S3}) and \eqref{Qs}, the corresponding asymptotic solutions can be obtained as
\begin{align}
	Q_{*,(\rm asy)}^{(\rm LL, NLL, N^{2}LL, N^{3}LL)}=\{46.0585, 55.1274, 55.2043, 55.1322\}\ {\rm GeV},
\end{align}
whose values are very close to Eq.~\eqref{Qs_ext} due to the excellent convergence, indicating the asymptotic solution \eqref{Qs} is a good approximation for the present case. In this work, the exact numerical solutions Eq.~\eqref{Qs_ext} are used.

According to the PMC conformal series Eq.~\eqref{PMC_Hbb} obtained in Sec.\ref{sec3}, using the N$^{3}$LL-accuracy PMC scale $Q_{*}$, we obtain the N$^{4}$LO-level pQCD approximant $\Gamma(H \to b\bar{b})$ under the PMC scale-setting approach
\begin{equation}
\Gamma(H \to b\bar{b})\big|_{\rm PMC}  = 2.3819\ {\rm MeV}, \label{numHbbPMC}
\end{equation}
in which the conventional scale dependence has been removed. On the other hand, the scale dependence of the original pQCD series diminishes with the inclusion of more loop terms, aligning with conventional wisdom. For the present N$^{4}$LO series, the net error under the conventional scale-setting approach is modest, amounting to approximately $0.09\%$ for $\mu_{r}\in [M_{H}/2, 2M_{H}]$, which increase to about $0.26\%$ for a wider scale interval $\mu_{r}\in [M_{H}/4, 4M_{H}]$. For clarity, we present the N$^{4}$LO-level conventional prediction as follows:
\begin{equation}
\Gamma(H \to b\bar{b})\big|_{\rm Conv.} = 2.3842_{-0.0020}^{+0.0002}\left(_{-0.0080}^{+0.0002}\right)\ {\rm MeV}, \label{numHbbConv}
\end{equation}
where the central value corresponds to $\mu_{r} = M_{H}$, the errors are for $\mu_{r}\in [M_{H}/2, 2M_{H}]$ (the errors in the parenthesis are for a wider range $\mu_{r}\in [M_{H}/4, 4M_{H}]$). The net N$^{4}$LO $\Gamma(H \to b\bar{b})$ for conventional and PMC series are consistent with each other under the proper choice of scale interval for the conventional scale-setting approach. Thus, we must be cautious when discussing the scale uncertainties within conventional scale-setting approach. For convenience, unless otherwise specified, we will adopt the usual choice of $\mu_{r}\in [M_{H}/2, 2M_{H}]$ for our discussions.

\begin{figure}[htb]
	\centering
	\includegraphics[width=0.88\textwidth]{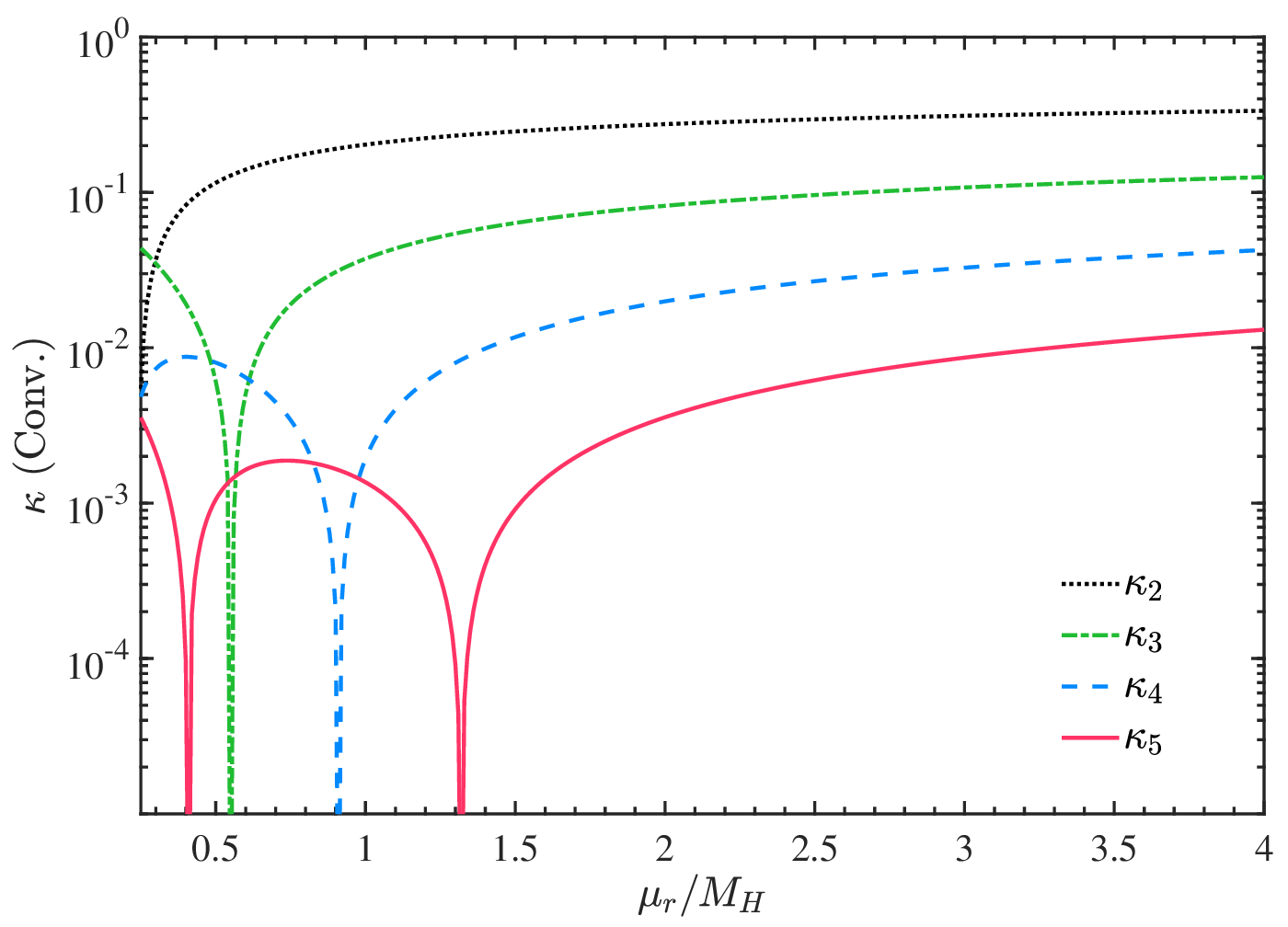}
	\caption{The $\kappa$-factor for conventional series versus the renormalization scale $\mu_{r}$. For convenience, the scale is non-dimensionalized by $M_{H}$. }
	\label{kappaConv}
\end{figure}

It is observed that the perturbative behavior of the conventional series still depends heavily on the choice of renormalization scale. To illustrate this point more clearly, we define a $\kappa$-factor for the N$^{4}$LO series (\ref{Conv_Hbb}) or (\ref{PMC_Hbb}), that is,
\begin{align}
\kappa_{i}(\rm Conv.) &= \left|\frac{r_{i}(\mu_{r})\alpha_{s}^{i-1}(\mu_{r})}{r_{1}}\right|, &
\kappa_{i}(\rm PMC)   &= \left|\frac{r_{i,0}\alpha_{s}^{i-1}(Q_{*})}{r_{1,0}}\right|,
\end{align}
where $\kappa_{i}(\rm Conv.)$ are highly scale-dependent, whose scale-running behaviors are shown in Fig.~\ref{kappaConv}. Specifically, we can compare $\kappa(\rm PMC)$ and $\kappa({\rm Conv.}) |_{\mu_{r} = M_{H}}$ numerically, e.g.
\begin{align}
	\kappa(\rm Conv.) &= \{1,0.2031,0.0374,0.0019,0.0014\},\quad \mu_{r} = M_{H}  \label{kappa1}  \\
	\kappa(\rm PMC)   &= \{1,0.0677,0.0067,0.0004,0.0001\}. \label{kappa2}
\end{align}
From the above expressions and Fig.~\ref{kappaConv}, it is evident that the convergence of PMC series is much better than that of conventional series. The scale-independent and more convergent PMC series~\eqref{PMC_Hbb} can be regarded as the intrinsic perturbative nature of $\tilde{R}(s)$.

\begin{figure}[htb]
	\centering
	\includegraphics[width=0.88\textwidth]{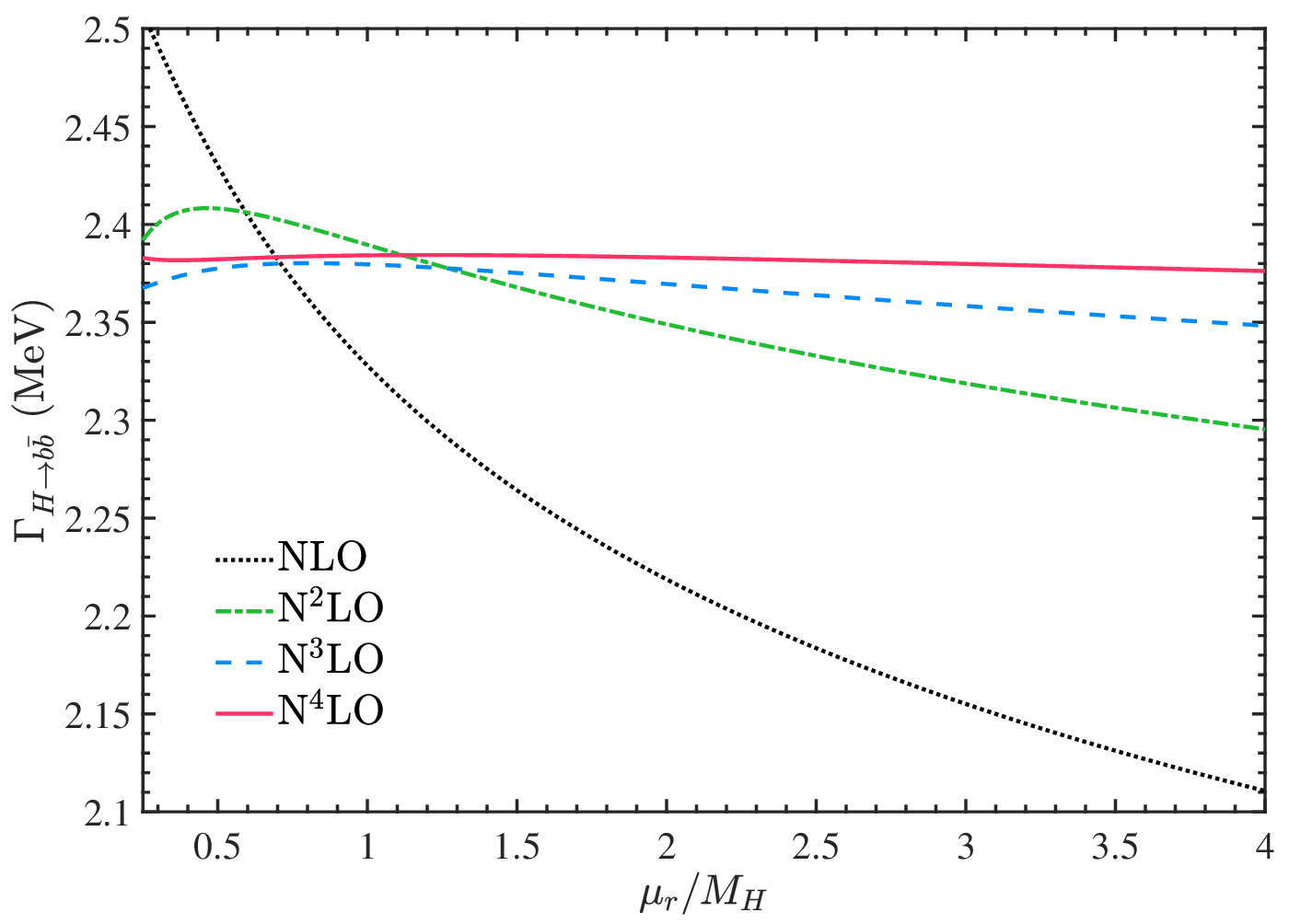}
	\includegraphics[width=0.88\textwidth]{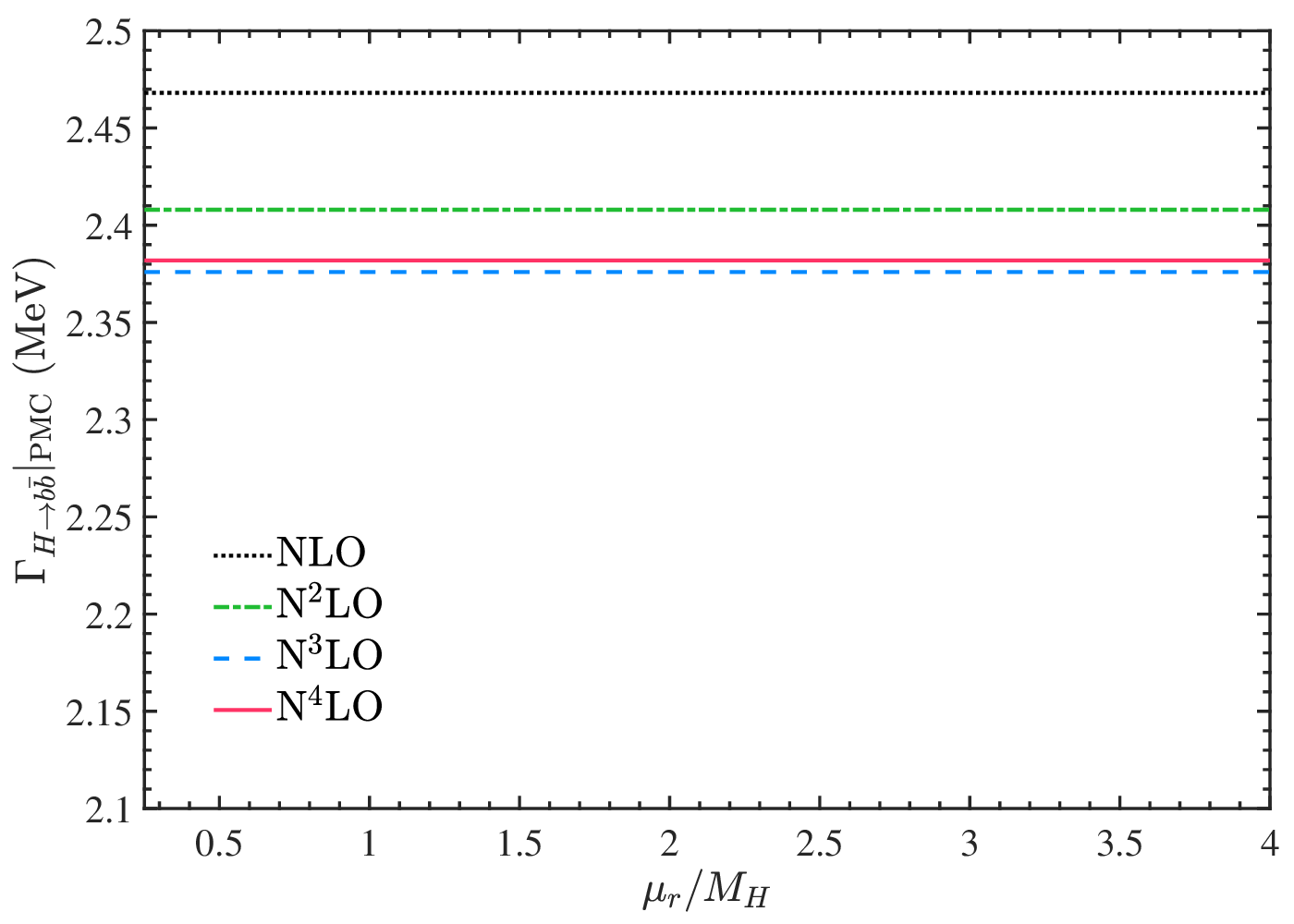}
	\caption{The total decay width $\Gamma(H\to b\bar{b})$ up to N$^{4}$LO QCD corrections versus the renormalization scale $\mu_{r}$ (divided by $M_{H}$) using conventional (Conv.) (upper) and PMC (lower) scale-setting approaches, respectively.}
	\label{Gamma_Hbb}
\end{figure}

We present the total decay width $\Gamma(H \to b\bar{b})$ up to N$^{4}$LO QCD corrections versus the renormalization scale $\mu_{r}$ (divided by $M_{H}$) before and after applying the PMC in Fig.~\ref{Gamma_Hbb}. After applying the PMC, as shown in Fig.~\ref{Gamma_Hbb}, the scale dependence of the total decay width is removed at any fixed order. The difference between N$^{3}$LO-level and the N$^{4}$LO-level PMC predictions is much smaller than that of the conventional ones, indicating that the convergence of the pQCD series is significantly improved.

\subsection{Predictions of the uncalculated N$^{5}$LO contributions using the Bayesian analysis approach}

As mentioned in Sec.~\ref{sec1}, a truncated perturbation series cannot not automatically satisfy the RGI requirements, leading to conventional renormalization scheme and scale ambiguities. These can be treated as one source of the error arising from the UHO-terms. The potential magnitude of the UHO-terms can be treated as another source of error in the fixed-order series, and how to achieve an unbiased estimation on it is also an important issue. Typically, the variation of perturbation series arising from changing the unphysical scales (e.g. renormalization and factorization scales) within a specified range is used to estimate both types of UHO errors. This straightforward treatment may delineate an order estimation of the UHO contributions pertaining only to the non-conformal terms that are linked to the scale evolution or equivalently the scale-dependent $\{\beta_i\}$-terms in the UHO-terms, which control the running of $\alpha_s$. Thus, a more nuanced and analytical framework that goes beyond mere scale considerations is demanded to systematically predict the UHO errors. Unfortunately, no such systematic method has been developed to date. In the following, we will adopt the Bayesian analysis (BA) method, being as a probabilistic interpretation framework, to estimate the possible contribution of the uncalculated N$^{5}$LO-terms of $\Gamma(H \to b\bar{b})$.

The BA method, as described in Refs.~\cite{Cacciari:2011ze, Bagnaschi:2014wea, Bonvini:2020xeo, Duhr:2021mfd}, provides a powerful tool to estimate theory uncertainties arising from the UHO-terms. The BA quantifies the contributions of the UHO-terms in terms of a probability distribution. Its effectiveness is maximized for a series with good convergent properties. A detailed explanation of BA procedures in combination with the PMC can be found in Ref.\cite{Shen:2022nyr}. In the following, we only present main results and the interested readers may turn to Ref.\cite{Shen:2022nyr} for detail. Following the idea of the BA, given a fixed degree-of-belief (DoB) or equivalently the Bayesian probability, the estimated UHO-coefficient $r_{p+1}$ based on the known coefficients $\{r_{1},r_{2},...,r_{p}\}$ will lie within a specific credible interval~(CI) of $r_{p+1}\in\left[-r_{p+1}^{(\rm DoB)}, r_{p+1}^{(\rm DoB)}\right]$, where
\begin{align}
	r_{p+1}^{(\rm DoB)}=\left\{
	\begin{aligned}
		&\bar{r}_{(p)}\frac{p+1}{p}{\rm DoB},&{\rm DoB}&\le \frac{p}{p+1}\\
		&\bar{r}_{(p)}\left[(p+1)(1-{\rm DoB})\right]^{-1/p},&{\rm DoB}&\ge \frac{p}{p+1}
	\end{aligned}\right.
\end{align}
with $\bar{r}_{(p)}={\rm max}\{|r_{1}|,|r_{2}|,...,|r_{p}|\}$. Therefore, the resulting predictive estimate for the perturbative quantity at the $(p+1)_{\rm th}$-order can be formally expressed as a CI centered at the known fixed-order result, with the uncertainty quantified through the predicted coefficient $r_{p+1}^{(\rm DoB)}$. Specifically, the next-order prediction $\rho_{p+1}$ for a general pQCD series truncated at the $p_{\rm th}$-order, $\rho_{p} = \sum_{i=1}^{p}r_{i}\alpha_{s}^{i}$, satisfies the following containment relation:
\begin{equation}\label{BA_CIs}
	\rho_{p+1} \in \left[\rho_{p}-r_{p+1}^{(\rm DoB)}\alpha_{s}^{p+1}, \rho_{p}+r_{p+1}^{(\rm DoB)}\alpha_{s}^{p+1}\right].
\end{equation}	
Here, the interval half-width $r_{p+1}^{(\rm DoB)}\alpha_{s}^{p+1}$ encodes the systematic uncertainty associated with the UHO-terms. This construction provides a probabilistic framework for estimating UHO corrections. For definiteness and without loss of generality, we take ${\rm DoB}\equiv 95.5\%$ to estimate the contributions from the UHO-terms.

It is noted that the improved series using the PMC scale-setting approach not only enhances the precision of the fixed-order pQCD series but also provides a solid foundation for estimating the potential contributions from the UHO-terms, thus significantly enhancing the predictive capabilities of perturbation theory. Some successful BA applications in conjunction with the PMC are documented in Refs.\cite{Du:2018dma, Shen:2022nyr, Yan:2022foz, Shen:2023qgz, Luo:2023cpa, Yan:2023mjj,Yan:2024hbz}. In the following, we will employ the BA method to estimate the contributions of the uncalculated N$^{5}$LO terms from the established initial N$^{4}$LO-level series~\eqref{Conv_Hbb} and the PMC series~\eqref{PMC_Hbb}.

\begin{figure} [htb]
\centering
\includegraphics[width=0.8\textwidth]{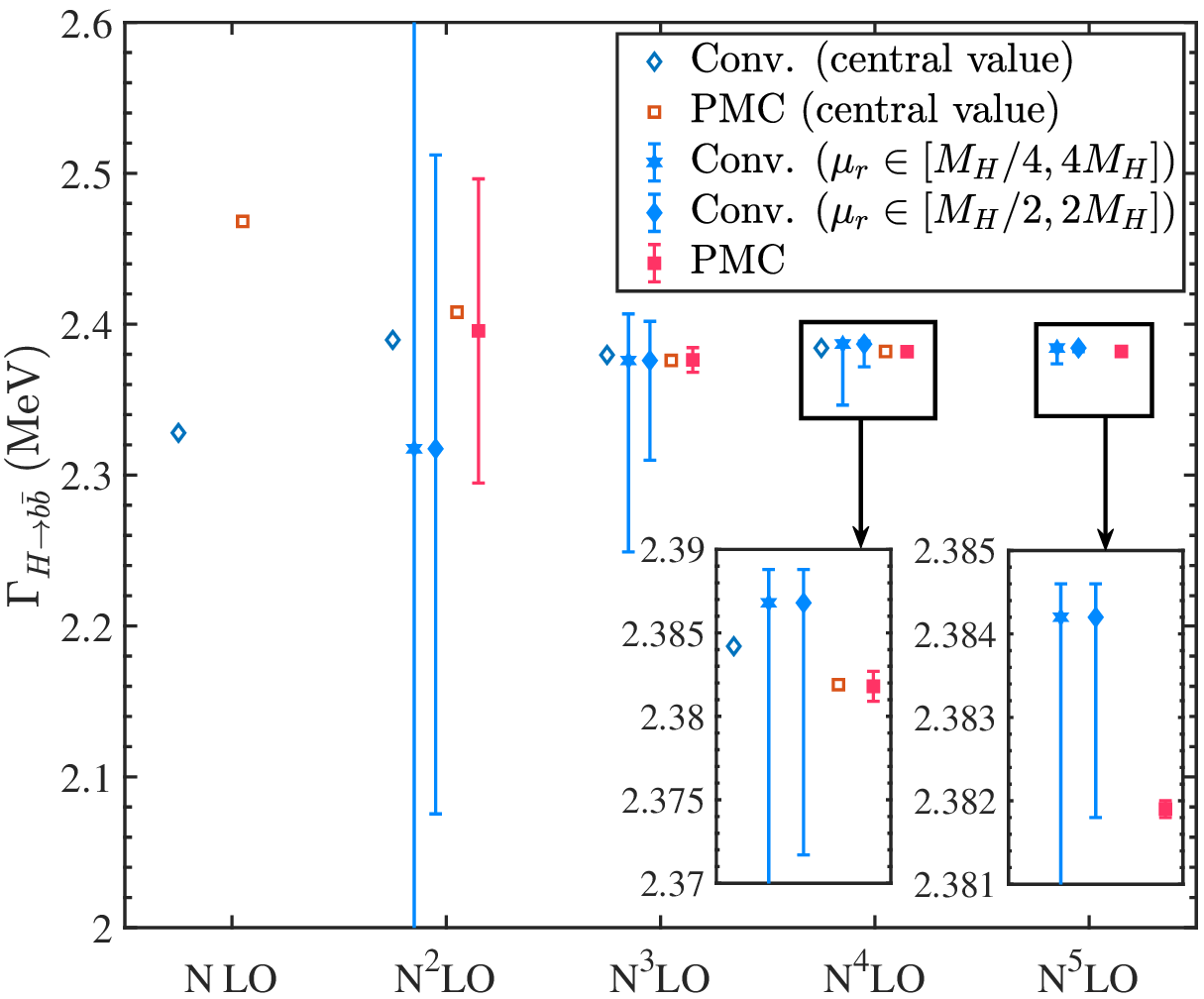}
\caption{Comparison of the calculated central values of the total decay width $\Gamma(H\to b\bar{b})$ for the known series (labeled as ``central value'') with the predicted credible intervals of $\Gamma(H \to b\bar{b})$ up to N$^{5}$LO-level QCD corrections. The blue hollow diamonds and red hollow squares represent the calculated central values of the fixed-order pQCD predictions using conventional (Conv.) and PMC scale-setting approaches, respectively. The blue stars, blue solid diamonds, and red solid squares with error bars represent the predicted credible intervals of the BA method based on the known Conv. series and the PMC series, respectively. ${\rm DoB}=95.5\%$. }
\label{UHOerr}
\end{figure}

The comparison between the central values calculated from known series (simply labeled as ``central value'') for the total decay width $\Gamma_{H\to b\bar{b}}$ with the predicted CIs of $\Gamma_{H\to b\bar{b}}$ up to N$^{5}$LO QCD corrections are presented in Fig.~\ref{UHOerr}. In Fig.~\ref{UHOerr}, the blue hollow diamonds and red hollow squares correspond to the numerically evaluated central values of the fixed-order pQCD series under conventional~(Conv.) and PMC scale-setting approaches, respectively. Additionally, the BA method-derived CIs, represented by the blue star, blue solid diamonds, and red solid squares with error bars, are generated by propagating the N$^{k-1}$LO Conv. and PMC series to predicted N$^{k}$LO results. Notably, while the kernel function $\tilde{R}$ alone would enforce equivalence between the N$^{k}$LO solid central values and the N$^{k-1}$LO hollow counterparts according to Eq.~\eqref{BA_CIs}, the inclusion of the full observable introduces an additional dependence on the $\overline{m}_{b}$ parameter. Crucially, the precision of $\overline{m}_{b}$ varies systematically with the perturbative order. This differential treatment of $\overline{m}_{b}$'s running across orders breaks the naive correspondence between solid and hollow central values, thereby accounting for their observed disparity. Due to the scale dependence of the coefficients $r_{i>1}$ in conventional pQCD series, the BA method can only be employed after specifying the renormalization scale, thus introducing additional uncertainties into the total decay width. In Fig.~\ref{UHOerr}, we present two predictions for $\mu_{r}\in [M_{H}/2, 2M_{H}]$ and $\mu_{r}\in [M_{H}/4, 4M_{H}]$, respectively. Conversely, the conformal coefficients $r_{i,0}$ in the PMC series are independent of scale, providing a more reliable foundation for constraining estimations from the UHO contributions. Figure~\ref{UHOerr} illustrates that the probability distributions become more accurate and the resulting CIs shrink for the same DoB as more loop terms are included. When using the BA method with a widely used scale range of $\mu_{r}$ within $[M_{H}/2, 2M_{H}]$ to estimate the uncertainties from the UHO-terms, the resulting errors are
\begin{align}
	\Delta\Gamma(H \to b\bar{b})\big|_{\rm Conv.}^{\rm UHO} &=\left(_{-0.0024}^{+0.0004}\right)\ {\rm MeV},\\
	\Delta\Gamma(H \to b\bar{b})\big|_{\rm PMC}^{\rm UHO}&= \pm0.0001\ {\rm MeV}.
\end{align}
Given the present known N$^4$LO-level QCD corrections, both conventional and PMC scale-setting approaches show the UHO-contributions are quite small, less than $0.1\%$ or $0.01\%$ for either series, respectively. This is due to the good convergent behavior of both series. Moreover, the predicted errors of the PMC series are scale independent and present  more quicker trends of approaching its ``physical/exact" value. And the predicted error range for the series under conventional scale-setting approach will be increased to a certain degree by choosing a broader range, e.g. $\Delta\Gamma(H \to b\bar{b})\big|_{\rm Conv.}^{\rm UHO} =\left(_{-0.0105}^{+0.0004}\right)\ {\rm MeV}$ for $\mu_{r}\in [M_{H}/4, 4M_{H}]$, in which the error increases to $\sim 0.5\%$.

\subsection{Uncertainties of the total decay width $\Gamma(H \to b\bar{b})$}

Besides the uncertainties associated the UHO-terms, there are other error sources in determining the total decay width $\Gamma(H\to b\bar{b})$. In the following, when analyzing the uncertainty from a particular error source, the other inputs will be fixed to be their central values.

\begin{table}[htbp]
\centering
\begin{tabular}{cccc}
\hline
		& ~~$\Delta\Gamma(H \to b\bar{b})\big|^{\Delta\alpha_{s}(M_{Z})}$~~ & ~~$\Delta\Gamma(H \to b\bar{b}) \big|^{\Delta\overline{m}_{b}(\overline{m}_{b})}$~~ & ~~$\Delta\Gamma(H \to b\bar{b})\big|^{\Delta M_{H}}$~~ \\
\hline
	~~Conv.~~ & $\left(_{-0.0210}^{+0.0208}\right)$ & $\left(_{-0.0094}^{+0.0095}\right)$ & $\pm0.0021$ \\
\hline
	~~PMC~~   & $\left(_{-0.0210}^{+0.0209}\right)$ & $\left(_{-0.0094}^{+0.0095}\right)$ & $\pm0.0017$ \\
\hline
\end{tabular}
\caption{Additional uncertainties (in unit: MeV) arising from $\Delta \alpha_{s}(M_{Z})=\pm 0.0009$, $\Delta \overline{m}_{b}(\overline{m}_{b}) = \pm 0.007$~GeV, and $\Delta M_{H}= \pm 0.11$~GeV under conventional and PMC scale-setting approaches, respectively. } \label{err}
\end{table}

We summarize the uncertainties due to $\Delta \alpha_{s}(M_{Z})=\pm 0.0009$, $\Delta \overline{m}_{b}(\overline{m}_{b}) = \pm 0.007$ GeV, and $\Delta M_{H}= \pm 0.11$ GeV for both the conventional and PMC scale-setting approaches in Table~\ref{err}. Table~\ref{err} indicates that $\Delta \alpha_{s}$ dominates the errors, with its impact on the decay width being approximately $2$ to $12$ times greater than that of other sources. The quadratic average of uncertainties for the Higgs boson decay width from $\Delta \alpha_{s}(M_{Z})$, $\Delta \overline{m}_{b}(\overline{m}_{b})$, $\Delta M_{H}$, and the predicted UHO-terms are
\begin{align}
	\Gamma(H \to b\bar{b})\big|_{\rm Conv.} &= 2.3842_{-0.0232}^{+0.0230} \ {\rm MeV}, \\
	\Gamma(H \to b\bar{b})\big|_{\rm PMC}   &= 2.3819_{-0.0231}^{+0.0230} \ {\rm MeV}.
\end{align}
Their uncertainties are close to each other since errors from $\Delta \alpha_{s}(M_{Z})$ are so dominant that other sorts of errors are heavily diluted. Consequently, a more precise determination of $\alpha_{s}$, including the magnitude of the reference point $\alpha_s(M_Z)$, will significantly improve the precision of theoretical predictions.

\section{Summary }  \label{sec5}

In the paper, we have defined the CO to rederive the scale-displacement relation, the QCD degeneracy relations, and the PMC formulas involving the running of $\overline{\rm MS}$ mass $\overline{m}_{q}$. This novel PMC formulas that are based on the CO, which can be conveniently extended to deal with other input quantities such as the parton distribution function that also evolve with the renormalization scale.

We have presented an improved analysis of the total decay width $\Gamma(H\to b\bar{b})$ up to N$^{4}$LO-level QCD corrections by applying the PMC CO formulas. In contrast to previous studies~\cite{Shen:2017pdu, Wang:2013bla}, our current treatment accounts for the running behaviors of both $\alpha_{s}$ and $\overline{m}_{b}$, simultaneously. As evidenced by Eqs.~\eqref{kappa1}-\eqref{kappa2} and Fig.~\ref{kappaConv}, the N$^{4}$LO-level QCD corrections to the total decay width $\Gamma(H \to b\bar{b})$ exhibit excellent pQCD convergence after applying the PMC.

The pQCD series has been enhanced by applying the PMC is independent of the choice of renormalization scale, resulting in a more precise prediction and providing a superior foundation for estimating the contributions of the UHO-terms. Figure~\ref{Gamma_Hbb} indicates that the differences between the N$^{3}$LO and the N$^{4}$LO PMC predictions are much smaller than the conventional one, demonstrating that the convergence of the pQCD series is significantly improved and the PMC prediction exhibits a faster trends of approaching its ``physical/exact" value. Consequently, our present results emphasize the necessity of employing appropriate scale-setting approaches to achieve precise fixed-order pQCD predictions. And this analysis also provides another example that the PMC scale-invariant series enhances the predictive power of the Bayesian-based approach. As a final remark, it is noted that the significant improvements on the perturbative property of the series achieved by applying the PMC are diluted by the error associated with $\Delta\alpha_{s}(M_{Z})$; this inversely can be adopted for fixing a more precise $\alpha_{s}(M_{Z})$, if $\Gamma(H\to b\bar{b})$ can be measured with high precision.

\hspace{1cm}

\section*{Acknowledgements}

Project supported by graduate research and innovation foundation of Chongqing, China (Grant No. CYB240057 and No.ydstd1912), the Natural Science Foundation of China under Grants No. 12175025 and No. 12347101, the Hunan Provincial Natural Science Foundation with Grant No. 2024JJ3004, and the YueLuShan Center for Industrial Innovation (\# 2024YCII0118).

\hspace{1cm}
\section*{Appdendix}

\appendix

\section{The expansion coefficients of ${\cal \hat{D}}_{n_{\gamma},n_{\beta}}^{k}[\alpha_{s}^{\ell}]$}\label{A1}

The expansion coefficients of ${\cal \hat{D}}_{n_{\gamma},n_{\beta}}^{k}[\alpha_{s}^{\ell}]$, which have been defined in Eq.~\eqref{series_Das}, can be derived via an order-by-order way. The required coefficients for the present N$^4$LO-level example are
\begin{align}
	d_{i}^{[n_{\gamma},n_{\beta};1,\ell]} =&\, n_{\gamma}\gamma_{i} + (n_{\beta}+\ell)\beta_{i},\label{di1}\\
	d_{0}^{[n_{\gamma},n_{\beta};2,\ell]} =&\, \left[n_{\gamma}\gamma_{0} + (n_{\beta}+\ell+1)\beta_{0}\right] \left[n_{\gamma}\gamma_{0} + (n_{\beta}+\ell)\beta_{0}\right],\label{d02}\\
	d_{1}^{[n_{\gamma},n_{\beta};2,\ell]} =&\, \left[n_{\gamma}\gamma_{0} + (n_{\beta}+\ell+2)\beta_{0}\right] \left[n_{\gamma}\gamma_{1} + (n_{\beta}+\ell)\beta_{1}\right]\notag\\
	&\, + \left[n_{\gamma}\gamma_{1} + (n_{\beta}+\ell+1)\beta_{1}\right] \left[n_{\gamma}\gamma_{0} + (n_{\beta}+\ell)\beta_{0}\right]\label{d12}\\
	d_{2}^{[n_{\gamma},n_{\beta};2,\ell]} =&\, \left[n_{\gamma}\gamma_{0} + (n_{\beta}+\ell+3)\beta_{0}\right] \left[n_{\gamma}\gamma_{2} + (n_{\beta}+\ell)\beta_{2}\right]\notag\\
	&\, + \left[n_{\gamma}\gamma_{1} + (n_{\beta}+\ell+2)\beta_{1}\right] \left[n_{\gamma}\gamma_{1} + (n_{\beta}+\ell)\beta_{1}\right]\notag\\
	&\, + \left[n_{\gamma}\gamma_{2} + (n_{\beta}+\ell+1)\beta_{2}\right] \left[n_{\gamma}\gamma_{0} + (n_{\beta}+\ell)\beta_{0}\right]\label{d22}\\
	d_{0}^{[n_{\gamma},n_{\beta};3,\ell]} =&\, \left[n_{\gamma}\gamma_{0} + (n_{\beta}+\ell+2)\beta_{0}\right] \left[n_{\gamma}\gamma_{0} + (n_{\beta}+\ell+1)\beta_{0}\right] \left[n_{\gamma}\gamma_{0} + (n_{\beta}+\ell)\beta_{0}\right]\label{d03}\\
	d_{1}^{[n_{\gamma},n_{\beta};3,\ell]} =&\, \left[n_{\gamma}\gamma_{0} + (n_{\beta}+\ell+3)\beta_{0}\right] \left[n_{\gamma}\gamma_{0} + (n_{\beta}+\ell+2)\beta_{0}\right] \left[n_{\gamma}\gamma_{1} + (n_{\beta}+\ell)\beta_{1}\right]\notag\\
	&\, + \left[n_{\gamma}\gamma_{0} + (n_{\beta}+\ell+3)\beta_{0}\right] \left[n_{\gamma}\gamma_{1} + (n_{\beta}+\ell+1)\beta_{1}\right] \left[n_{\gamma}\gamma_{0} + (n_{\beta}+\ell)\beta_{0}\right]\notag\\
	&\, + \left[n_{\gamma}\gamma_{1} + (n_{\beta}+\ell+2)\beta_{1}\right] \left[n_{\gamma}\gamma_{0} + (n_{\beta}+\ell+1)\beta_{0}\right] \left[n_{\gamma}\gamma_{0} + (n_{\beta}+\ell)\beta_{0}\right]\label{d13}\\
	d_{0}^{[n_{\gamma},n_{\beta};4,\ell]} =&\ \left[n_{\gamma}\gamma_{0} + (n_{\beta}+\ell+3)\beta_{0}\right] \left[n_{\gamma}\gamma_{0} + (n_{\beta}+\ell+2)\beta_{0}\right] \left[n_{\gamma}\gamma_{0} + (n_{\beta}+\ell+1)\beta_{0}\right]\notag\\
	&\,\times \left[n_{\gamma}\gamma_{0} + (n_{\beta}+\ell)\beta_{0}\right]\label{d04}.
\end{align}

\section{The coefficients of asymptotic solution $\ln \left(Q_{*}^{2}/Q^{2}\right)$ for PMC scale}\label{A2}

The perturbative coefficients of the asymptotic solution $\ln \left(Q_{*}^{2}/Q^{2}\right)$ in Eq.~\eqref{Qs} are
\begin{align}
	S_{0} =&\, -\frac{\hat{r}_{2,1}}{\hat{r}_{1,0}},\label{S0}\\
	S_{1} =&\, -\frac{\hat{r}_{2,1}}{\hat{r}_{1,0}} \left[ \frac{d_{0}^{[n_{\gamma},n_{\beta};1,1]}}{d_{0}^{[n_{\gamma},n_{\beta};1,0]}} \left( \frac{\hat{r}_{3,1}}{\hat{r}_{2,1}} - \frac{\hat{r}_{2,0}}{\hat{r}_{1,0}} \right) + \frac{1}{2!} \frac{d_{0}^{[n_{\gamma},n_{\beta};2,0]}}{d_{0}^{[n_{\gamma},n_{\beta};1,0]}} \left( \frac{\hat{r}_{3,2}}{\hat{r}_{2,1}} - \frac{\hat{r}_{2,1}}{\hat{r}_{1,0}} \right) \right],\label{S1}\\
	S_{2} =&\, -\frac{\hat{r}_{2,1}}{\hat{r}_{1,0}} \bigg[ \frac{d_{1}^{[n_{\gamma},n_{\beta};1,1]}}{d_{0}^{[n_{\gamma},n_{\beta};1,0]}} \left( \frac{\hat{r}_{3,1}}{\hat{r}_{2,1}} - \frac{\hat{r}_{2,0}}{\hat{r}_{1,0}} \right) + \frac{d_{0}^{[n_{\gamma},n_{\beta};1,2]}}{d_{0}^{[n_{\gamma},n_{\beta};1,0]}} \left( \frac{\hat{r}_{4,1}}{\hat{r}_{2,1}} - \frac{\hat{r}_{3,0}}{\hat{r}_{1,0}} \right)\notag\\
	&\, + \frac{1}{2!} \frac{d_{1}^{[n_{\gamma},n_{\beta};2,0]}}{d_{0}^{[n_{\gamma},n_{\beta};1,0]}} \left( \frac{\hat{r}_{3,2}}{\hat{r}_{2,1}} - \frac{\hat{r}_{2,1}}{\hat{r}_{1,0}} \right) + \frac{1}{2!} \frac{d_{0}^{[n_{\gamma},n_{\beta};2,1]}}{d_{0}^{[n_{\gamma},n_{\beta};1,0]}} \left( \frac{\hat{r}_{4,2}}{\hat{r}_{2,1}} - \frac{\hat{r}_{2,1}}{\hat{r}_{1,0}} \frac{\hat{r}_{2,0}}{\hat{r}_{1,0}} \right) \notag\\
	&\, + \frac{1}{3!} \frac{d_{0}^{[n_{\gamma},n_{\beta};3,0]}}{d_{0}^{[n_{\gamma},n_{\beta};1,0]}} \left( \frac{\hat{r}_{4,3}}{\hat{r}_{2,1}} - \frac{\hat{r}_{2,1}^{2}}{\hat{r}_{1,0}^{2}} \right) \bigg] - \left( \frac{d_{1}^{[n_{\gamma},n_{\beta};1,0]}}{d_{0}^{[n_{\gamma},n_{\beta};1,0]}} + \frac{d_{0}^{[n_{\gamma},n_{\beta};1,1]}}{d_{0}^{[n_{\gamma},n_{\beta};1,0]}} \frac{\hat{r}_{2,0}}{\hat{r}_{1,0}} + \frac{d_{0}^{[n_{\gamma},n_{\beta};2,0]}}{d_{0}^{[n_{\gamma},n_{\beta};1,0]}} \frac{\hat{r}_{2,1}}{\hat{r}_{1,0}} \right)S_{1}\notag\\
	&\, - \beta_{0} \frac{\hat{r}_{2,1}}{\hat{r}_{1,0}}S_{1},\label{S2}\\
	S_{3} =&\, -\frac{\hat{r}_{2,1}}{\hat{r}_{1,0}} \bigg[ \frac{d_{2}^{[n_{\gamma},n_{\beta};1,1]}}{d_{0}^{[n_{\gamma},n_{\beta};1,0]}} \left( \frac{\hat{r}_{3,1}}{\hat{r}_{2,1}} - \frac{\hat{r}_{2,0}}{\hat{r}_{1,0}} \right) + \frac{d_{1}^{[n_{\gamma},n_{\beta};1,2]}}{d_{0}^{[n_{\gamma},n_{\beta};1,0]}} \left( \frac{\hat{r}_{4,1}}{\hat{r}_{2,1}} - \frac{\hat{r}_{3,0}}{\hat{r}_{1,0}} \right) + \frac{d_{0}^{[n_{\gamma},n_{\beta};1,3]}}{d_{0}^{[n_{\gamma},n_{\beta};1,0]}} \left( \frac{\hat{r}_{5,1}}{\hat{r}_{2,1}} - \frac{\hat{r}_{4,0}}{\hat{r}_{1,0}} \right) \notag\\
	&\, + \frac{1}{2!} \frac{d_{2}^{[n_{\gamma},n_{\beta};2,0]}}{d_{0}^{[n_{\gamma},n_{\beta};1,0]}} \left( \frac{\hat{r}_{3,2}}{\hat{r}_{2,1}} - \frac{\hat{r}_{2,1}}{\hat{r}_{1,0}} \right) + \frac{1}{2!} \frac{d_{1}^{[n_{\gamma},n_{\beta};2,1]}}{d_{0}^{[n_{\gamma},n_{\beta};1,0]}} \left( \frac{\hat{r}_{4,2}}{\hat{r}_{2,1}} - \frac{\hat{r}_{2,1}}{\hat{r}_{1,0}} \frac{\hat{r}_{2,0}}{\hat{r}_{1,0}} \right)\notag\\
	&\, + \frac{1}{2!} \frac{d_{0}^{[n_{\gamma},n_{\beta};2,2]}}{d_{0}^{[n_{\gamma},n_{\beta};1,0]}} \left( \frac{\hat{r}_{5,2}}{\hat{r}_{2,1}} - \frac{\hat{r}_{2,1}}{\hat{r}_{1,0}} \frac{\hat{r}_{3,0}}{\hat{r}_{1,0}} \right) + \frac{1}{3!} \frac{d_{1}^{[n_{\gamma},n_{\beta};3,0]}}{d_{0}^{[n_{\gamma},n_{\beta};1,0]}} \left( \frac{\hat{r}_{4,3}}{\hat{r}_{2,1}} - \frac{\hat{r}_{2,1}^{2}}{\hat{r}_{1,0}^{2}} \right)\notag\\
	&\, + \frac{1}{3!} \frac{d_{0}^{[n_{\gamma},n_{\beta};3,1]}}{d_{0}^{[n_{\gamma},n_{\beta};1,0]}} \left( \frac{\hat{r}_{5,3}}{\hat{r}_{2,1}} - \frac{\hat{r}_{2,1}^{2}}{\hat{r}_{1,0}^{2}} \frac{\hat{r}_{2,0}}{\hat{r}_{1,0}} \right) + \frac{1}{4!} \frac{d_{0}^{[n_{\gamma},n_{\beta};4,0]}}{d_{0}^{[n_{\gamma},n_{\beta};1,0]}} \left( \frac{\hat{r}_{5,4}}{\hat{r}_{2,1}} - \frac{\hat{r}_{2,1}^{3}}{\hat{r}_{1,0}^{3}} \right) \bigg]\notag\\
	&\, - \bigg[ \frac{d_{2}^{[n_{\gamma},n_{\beta};1,0]}}{d_{0}^{[n_{\gamma},n_{\beta};1,0]}} + \frac{d_{1}^{[n_{\gamma},n_{\beta};1,1]}}{d_{0}^{[n_{\gamma},n_{\beta};1,0]}} \frac{\hat{r}_{2,0}}{\hat{r}_{1,0}} + \frac{d_{0}^{[n_{\gamma},n_{\beta};1,2]}}{d_{0}^{[n_{\gamma},n_{\beta};1,0]}} \frac{\hat{r}_{3,0}}{\hat{r}_{1,0}} + \frac{\hat{r}_{2,1}}{\hat{r}_{1,0}} \bigg( \frac{d_{1}^{[n_{\gamma},n_{\beta};2,0]}}{d_{0}^{[n_{\gamma},n_{\beta};1,0]}} + \frac{d_{0}^{[n_{\gamma},n_{\beta};2,1]}}{d_{0}^{[n_{\gamma},n_{\beta};1,0]}} \frac{\hat{r}_{2,0}}{\hat{r}_{1,0}}\notag\\
	&\, + \frac{1}{2!} \frac{d_{0}^{[n_{\gamma},n_{\beta};3,0]}}{d_{0}^{[n_{\gamma},n_{\beta};1,0]}} \frac{\hat{r}_{2,1}}{\hat{r}_{1,0}} \bigg) \bigg] S_{1} - \left( \frac{d_{1}^{[n_{\gamma},n_{\beta};1,0]}}{d_{0}^{[n_{\gamma},n_{\beta};1,0]}} + \frac{d_{0}^{[n_{\gamma},n_{\beta};1,1]}}{d_{0}^{[n_{\gamma},n_{\beta};1,0]}} \frac{\hat{r}_{2,0}}{\hat{r}_{1,0}} + \frac{d_{0}^{[n_{\gamma},n_{\beta};2,0]}}{d_{0}^{[n_{\gamma},n_{\beta};1,0]}} \frac{\hat{r}_{2,1}}{\hat{r}_{1,0}} \right) \left( S_{2} + \beta_{0} \frac{\hat{r}_{2,1}}{\hat{r}_{1,0}} S_{1} \right)\notag\\
	&\, + \frac{1}{2!} \frac{d_{0}^{[n_{\gamma},n_{\beta};2,0]}}{d_{0}^{[n_{\gamma},n_{\beta};1,0]}} S_{1}^{2} + \beta_{0} S_{1}^{2} - \beta_{1} \frac{\hat{r}_{2,1}}{\hat{r}_{1,0}} S_{1} - \beta_{0}^{2} \frac{\hat{r}_{2,1}^{2}}{\hat{r}_{1,0}^{2}} S_{1} - 2 \beta_{0} \frac{\hat{r}_{2,1}}{\hat{r}_{1,0}} S_{2},\label{S3}
\end{align}
which are consistent with those in Ref.~\cite{Huang:2022rij}.

\section{The coefficients of the correlator $\tilde{\Pi}(\mu;Q)/Q^{2}$ and the anomalous dimension $\gamma^{\rm SS}$} \label{A3}

The coefficients of correlator at the scale $\mu = Q$ are listed, where all logarithms vanish. Results for arbitrary values of $\mu$ can be recovered by solving the RGE \eqref{RGE_Pi}.
\begin{align}
	\Pi_{0} =&\, -4,\\
	\Pi_{1} =&\, \frac{1}{\pi} C_{F} \left( -\frac{131}{8} + 6 \zeta_{3} \right),\\
	\Pi_{2} =&\, \frac{1}{\pi^{2}} C_{F} \bigg[ C_{F} \left( -\frac{1613}{64} + 24 \zeta_{3} - \frac{9}{4} \zeta_{4} - 15 \zeta_{5} \right) + C_{A}\left( -\frac{14419}{288} + \frac{75}{4} \zeta_{3} +\frac{9}{8} \zeta_{4} + \frac{5}{2} \zeta_{5} \right)\notag\\
	&\, + T_{F} n_{f} \left( \frac{511}{36} - 4\zeta_{3} \right) \bigg],\\
	\Pi_{3} =&\, -\frac{4}{\pi^{3}} \bigg[ \left( \frac{215626549}{248832} - \frac{1789009}{3456}\zeta_{3} + \frac{1639}{32} \zeta_{3}^{2} - \frac{1645}{1152} \zeta_{4} + \frac{73565}{1728} \zeta_{5} + \frac{325}{192} \zeta_{6} - \frac{665}{72} \zeta_{7} \right)\notag\\
	&\, + n_{f} \left( -\frac{26364175}{373248} + \frac{22769}{864} \zeta_{3} - \frac{5}{6} \zeta_{3}^{2} - \frac{53}{48} \zeta_{4} + \frac{1889}{432} \zeta_{5} \right)\notag\\
	&\, + n_{f}^{2} \left( \frac{499069}{559872} - \frac{157}{1296} \zeta_{3} + \frac{1}{48} \zeta_{4} - \frac{5}{18}\zeta_{5} \right) \bigg],
\end{align}
where $\zeta_{n} \equiv \sum_{k=1}^{\infty} k^{-n}$ is Riemann's Zeta function. The coefficients of the anomalous dimension $\gamma^{\rm SS}$ are
\begin{align}
	\gamma^{\rm SS}_{0} =&\, -2,\\
	\gamma^{\rm SS}_{1} =&\, \frac{1}{\pi} C_{F} \left(-\frac{5}{2}\right) \\
	\gamma^{\rm SS}_{2} =&\, \frac{1}{\pi^{2}} C_{F} \left[ C_{F} \left( \frac{119}{32} - \frac{9}{2} \zeta_{3} \right) + C_{A} \left( -\frac{77}{16} + \frac{9}{4} \zeta_{3} \right) + T_{F} n_{f} \right],\\
	\gamma^{\rm SS}_{3} =&\, \frac{1}{\pi^{3}} C_{F} \bigg[ C_{F}^{2} \left( -\frac{4651}{384} - \frac{29}{4} \zeta_{3} + \frac{27}{8} \zeta_{4} + \frac{45}{4} \zeta_{5} \right) + C_{F}C_{A} \left( \frac{641}{48} - \frac{259}{16} \zeta_{3} +\frac{39}{16} \zeta_{4} + \frac{45}{8} \zeta_{5} \right)\notag\\
	&\, + C_{A}^{2} \left( -\frac{267889}{31104} + \frac{475}{48} \zeta_{3} - \frac{33}{16} \zeta_{4} - \frac{45}{8} \zeta_{5} \right) + C_{F} T_{F} n_{f} \left( \frac{125}{32} - \frac{1}{2} \zeta_{3} - 3 \zeta_{4} \right)\notag\\
	&\, + C_{A} T_{F} n_{f} \left( \frac{631}{7776} + \frac{5}{3} \zeta_{3} + \frac{9}{4} \zeta_{4} \right) + T_{F}^{2}n_{f}^{2} \left( \frac{1625}{1944} -\frac{2}{3} \zeta_{3} \right) \bigg],\\
	\gamma^{\rm SS}_{4} =&\, \frac{1}{3\pi^{4}}\bigg[ -\frac{1305623}{864} - \frac{540883}{3456} \zeta_{3} - \frac{19327}{288}\zeta_{3}^{2} - \frac{113557}{384} \zeta_{4} + \frac{158765}{576}\zeta_{5} + \frac{29825}{64} \zeta_{6} + \frac{97895}{384} \zeta_{7} \notag\\
	&\, + n_{f} \left( \frac{11341807}{62208} + \frac{385147}{1728} \zeta_{3} - \frac{187}{16} \zeta_{3}^{2} + \frac{10207}{192} \zeta_{4} - \frac{55127}{288} \zeta_{5} - \frac{6725}{96} \zeta_{6} \right)\notag\\
	&\, + n_{f}^{2} \left( \frac{249113}{373248} - \frac{749}{48} \zeta_{3} + \frac{21}{8} \zeta_{4} + \frac{37}{4} \zeta_{5} \right) + n_{f}^{3} \left( \frac{1625}{15552} + \frac{5}{108} \zeta_{3} - \frac{1}{6} \zeta_{4} \right) \bigg].
\end{align}

\section{An explanation on the non-absorption of $n_{f}$-terms in $\gamma^{\rm SS}$} \label{A4}

Practically, in doing loop calculation, we always get the $n_{f}$-series but not the $\{\beta_{i}\}$-series, and then in order to fix the correct magnitude of $\alpha_{s}$, we have to transfer the $n_{f}$-series into the required $\{\beta_{i}\}$-series. In addition to PMC, such a transformation has also been done by other groups, such as Refs.~\cite{Kataev:2014jba, Garkusha:2018mua, Kataev:2023xqr, Baikov:2022zvq, Mikhailov:2024mrs}. The conclusion drawn by Kataev and his collaborators using those transformations~\cite{Kataev:2014jba, Garkusha:2018mua, Kataev:2023xqr}, e.g. ``the correct $\beta$-expansion of the $D$-function includes also the $n_f$-terms of the photon anomalous dimension ($\gamma$), and all of them should be applied for PMC", should be accepted with care. 

The PMC determines the correct magnitude of the effective coupling constant of the process with the help of RGE. By using the PMC, the effective coupling constant is fixed by requiring that all the RGE-involved $\{\beta_{i}\}$-terms be eliminated. Thus, when applying the PMC, all $n_{f}$-terms must be separated and related to their own scale-running parameters such as the $\alpha_{s}$, or the quark masses, or the composite operators, and etc.; especially, only the RGE-involved $n_{f}$-terms need to be adopted for fixing the magnitude of $\alpha_{s}$. We take the non-singlet Adler function $D^{\rm ns}$ as an example to show why the $n_{f}$-terms in the anomalous dimension cannot be adopted for fixing the magnitude of $\alpha_s$. It is defined as~\cite{Baikov:2012zm}
\begin{eqnarray}
	D^{\rm ns}(a_s) = -12{\pi^2}{Q^2}\frac{\rm d}{{\rm d}{Q^2}}\Pi^{\rm ns}(L,{a_s}),
	\label{eq:Adler}
\end{eqnarray}
where $L=\ln{\mu^2}/{Q^2}$, $\mu$ encodes the renormalization scale. The coefficient $\Pi^{\rm ns}(L,{a_s})$ is the non-singlet part of the polarization function for a flavor-singlet vector current. The running behavior of $\Pi^{\rm ns}(L,a_s)$ is controlled by
\begin{eqnarray}
	\left( {\mu^2}\frac{\partial}{\partial{\mu^2}} + \beta(a_s)\frac{\partial}{\partial{a_s}} \right)\Pi^{\rm ns}(L,a_s) &=& \gamma^{\rm ns}(a_s) ,
	\label{eq:gamma}
\end{eqnarray}
where $\Pi^{\rm ns}(L,a_s)=\sum\limits_{i \ge 0} {\Pi_i^{\rm ns}} {a_s^i}/{16\pi^2}$. The Adler function depends on the anomalous dimension of the photon field $\gamma^{\rm ns}(a_s) = \sum\limits_{i \ge 0} {\gamma^{\rm ns}_i} {a_s^i}/{16\pi^2}$. Thus we obtain
\begin{equation}
	D^{\rm ns}(a_s) = 12\pi^2 \left[\gamma^{\rm ns}(a_s) - \beta(a_s)\frac{\partial} {\partial{a_s}}{\Pi}^{\rm ns}(L,{a_s})\right]. \label{eq:Dexpression}
\end{equation}
As a combination of all those equations, we finally have
\begin{equation}
	\mu^{2}\frac{{\rm d} D^{\rm ns}}{d\mu^2}=0
\end{equation}
at any fixed order, the pQCD approximant $D^{\rm ns}$ is thus a local RGI quantity. It should be emphasized that the anomalous dimension $\gamma^{\rm ns}(a_s)$ is associated with the renormalization of the QED coupling, it determines the correct running behavior of $\Pi^{\rm ns}(L,a_s)$; and it ensures that $D^{\rm ns}$ satisfy the local RGI, but not the standard RGI~\cite{Wu:2014iba}. This explains why the QED anomalous dimension $\gamma^{\rm ns}$, which appears in the definition of the Adler function, cannot be used to set the correct magnitude of the effective $\alpha_s$ for $D^{\rm ns}$.

In fact, if the $\{\beta_i\}$-terms in anomalous dimensions have been incorrectly included in the PMC treatment to obtain the $\beta$-pattern of the series, wrong PMC prediction for $R_{e^+ e^-}$ will appear, as has been done by Ref.\cite{Kataev:2014jba}. Subsequently, such misuse of PMC has been clarified by Ref.\cite{Ma:2015dxa}, which shows that if one treats anomalous dimensions correctly, one can obtain correct PMC predictions. And lately, it has been shown that accurate generalized scheme-independent Crewther relation can also be obtained by correctly treating anomalous dimensions~\cite{Shen:2016dnq, Huang:2020gic}; and etc.. 

As for the present case, the anomalous dimension $\gamma^{\rm SS}$ is related to the polarization function $\tilde{\Pi}$, which is analogous to the anomalous dimensions of the masses, thus $n_{f}$-terms related to it must also be considered as responsible of running of the particular operator/polarization function but not be responsible for the running of the coupling $\alpha_{s}$. This is why we have not considered the $n_{f}$-terms of the anomalous dimension $\gamma^{\rm SS}$. Moreover, in the definitions~\eqref{Adler1} and \eqref{Adler2} of the Adler function $\tilde{D}$, the contribution of the anomalous dimension $\gamma^{\rm SS}$ related to the polarization function is subtracted out leaving the Adler function $\overline{m}_{b}^{2}\tilde{D}$ scale invariant. Thus, if one reabsorbs the $n_{f}$-terms of the anomalous dimension into the Adler function, it would break its scale invariance.

To sum up, our point of view is that, ``one can transform the $n_{f}$-terms in anomalous dimensions also into $\{\beta_{i}\}$-terms, however one should handle those terms with caution to obtain correct magnitude of $\alpha_s$ and then correct PMC predictions". It is also noted that those $n_{f}$-terms in anomalous dimensions may have their own importance, which also need to be transformed into a separate $\beta$-series~\cite{futurework}; For example, the $n_f$-terms in the photon anomalous dimension of $D^{\rm ns}$ can be adopted for fixing the magnitude of the QCD-induced QED coupling constant via its own RGE that is given in Eq.(4.2) of Ref.\cite{Baikov:2012zm}.

\end{document}